\documentclass[11pt]{article}
\usepackage{amsfonts,amssymb,amsmath,epsfig,graphicx}
\usepackage{theorem,calc}
\usepackage{graphicx}
\usepackage{caption}
\usepackage{subcaption}
\usepackage{cite}
\usepackage{multicol}
\usepackage{vmargin}
\usepackage{fancyhdr}
\usepackage{url}

\usepackage{accents}

\usepackage{tikz}
\usetikzlibrary{arrows}
\tikzstyle{block}=[draw opacity=0.7,line width=1.4cm]
\usetikzlibrary{positioning,arrows,chains,matrix,scopes,fit,decorations.markings,decorations}

\newtheorem{definition}{Definition}[section]

\newtheorem{proposition}[definition]{Proposition}

\newtheorem{lemma}[definition]{Lemma}
\numberwithin{equation}{section}

\newtheorem{rmk}{Remark}[section]

\newcommand{\beq}{\begin{equation}}
\newcommand{\eeq}{\end{equation}}
\newcommand{\bea}{\begin{eqnarray}}
\newcommand{\eea}{\end{eqnarray}}
\newcommand{\beano}{\begin{eqnarray*}}
\newcommand{\eeano}{\end{eqnarray*}}
\newcommand{\bma}{\begin{pmatrix}}
\newcommand{\ema}{\end{pmatrix}}

\newcommand{\CC}{{\mathbb C}}

\newcommand{\RR}{\mbox{${\mathbb R}$}}

\newcommand{\prf}{\underline{Proof:}\ }
\newcommand{\finprf}{\null
  \hfill {\rule{5pt}{5pt}}} 
\newcommand{\ie}{{\it i.e.}\ }

\DeclareMathOperator{\tr}{tr}

\title{Dressing the boundary: on soliton solutions of  \\  the nonlinear Schr\"odinger equation  on the half-line }
\author{  Cheng Zhang \\ \url{ch.zhang.maths@gmail.com} \\ \\
 Department of Mathematics, Shanghai University\\
99 Shangda Road, Baoshan District\\
Shanghai, 200444, China}
\date{\empty}
\makeatletter
\newcommand\appendix@section[1]{%
\refstepcounter{section}%
\orig@section*{Appendix \@Alph\c@section: #1}%
\addcontentsline{toc}{section}{Appendix \@Alph\c@section: #1}%
}
\let\orig@section\section
\g@addto@macro\appendix{\let\section\appendix@section}
\makeatother

\begin{document}
\maketitle
\begin{abstract}
  Based on the theory of integrable boundary conditions (BCs) developed by Sklyanin, we provide a direct method for computing soliton solutions of the focusing nonlinear Schr\"odinger (NLS) equation on the half-line.  The integrable BCs at the origin are represented by constraints of the Lax pair, and our method lies on dressing the Lax pair by preserving those constraints in the Darboux-dressing process. The method is applied to two classes of solutions: solitons vanishing at infinity and self-modulated solitons on a constant background. Half-line solitons in both cases are explicitly computed. In particular, the boundary-bound solitons, that are static solitons bounded at the origin, are also constructed. We give a natural inverse scattering transform interpretation of the method as evolution of the scattering data determined by the integrable BCs in space. 
\vspace{.2cm}




\end{abstract}

\section{Introduction}
Research in initial-boundary-value problems for integrable nonlinear PDEs represents one of the basic problems in integrable systems. One crucial aspect is that the boundary conditions (BCs) are inherent in the definition of integrability. For instance, in the inverse scattering transform (IST) approach to integrating the Korteweg-de Vries (KdV) equation, the vanishing BCs (the KdV field and its derivatives vanish as the space variable tends to infinity) are explicitly imposed \cite{kdv}. In the Hamiltonian formulation of integrable field theory, the vanishing BCs  are also needed in order to ensure the existence of infinitely many conserved quantities \cite{G1,f1}. Indeed, one could argue that integrable PDEs are said to be integrable only if they obey certain BCs that  preserve the integrability of the system. The common choices are vanishing BCs or  periodic BCs as for the case of KdV.

One of the systematic approaches to boundary-value problems to nonlinear PDEs were due to Sklyanin \cite{sklyanin1987boundary} in the framework of the Hamiltonian field theory, cf.~\cite{faddeev}. In Sklyanin's approach, the central object is  the so-called classical reflection equation involving the classical $r$ matrix and boundary $K_\pm$ matrices. The BCs at the two ends of an interval are encoded into $K_\pm$. Then the notion of integrability in the presence of BCs is clearly defined as the existence of infinitely many commuting flows. In fact,  the theory is related to a far-reaching context as it represents the classical aspects of the quantum theory of integrability  \cite{SKBC, Cherednik1}.

The IST is the analytic approach for solving initial-value problems for integrable PDEs, cf.~\cite{AKNS, faddeev}. To address to integrable PDEs on the half-line, or more generally on an interval with generic BCs, the IST has been remarkably generalized by Fokas to the unified transform method \cite{fokas1997unified, fokas2002integrable, fokas2008}. The key idea of the unified transform method is lying on the simultaneous treatments of both the initial and boundary data in the direct scattering process. Then the scattering data are put into certain functional constraints usually formulated as Riemann-Hilbert problems in the inverse space.

Although Fokas' approach has the great advantage that both the initial and boundary data are regarded on an equal footing, in general, it is a difficult task to solve the inverse (Riemann-Hilbert) problems to derive explicit solutions of the original PDEs. Moreover, in contrast to Sklyanin's approach, there is no clear definition of integrable BCs in the unified transform method. Note that a special class of BCs, known as {\em linearizable} BCs, do exist in Fokas' approach \cite{fokas2002integrable, fokas2008}. The linearizable BCs reflect certain symmetry of the scattering data that can be used to simplify the Riemann-Hilbert problems. For certain models, they coincide with the integrable BCs, cf.~\cite{sklyanin1987boundary, fokas2002integrable}. However, linearizable BCs are, a priori, not equivalent to integrable BCs (see Section $4$ for a detailed example).  
  As an important application of the unified transform method, asymptotic solutions of integrable PDEs at large time, mostly accompanied with the linearizable BCs, can be derived \cite{fokas2002integrable, fokas2008}.  

 In this paper, we study the focusing nonlinear Schr\"odinger (NLS) equation 
\begin{equation}
  \label{eq:nls}
  i q_t + q_{xx}-2  |q|^2 q= 0 \,,\quad q:=q(x,t)\,,
\end{equation}
restricted to the half-line domain, \ie $x\geq 0$, under the Robin BCs at $x=0$. It is known that the Robin BCs are integrable BCs \cite{sklyanin1987boundary}, and our aim is to compute exact solutions of the NLS equation on the half-line in imposing the integrable BCs. 

The NLS equation is an important model in mathematical physics (see for instance \cite{faddeev, ZS2, APT} for its integrable aspects). It can be solved by the IST  under the vanishing BCs \cite{ZS2}, or on a constant background that has non-vanishing BCs \cite{ZS3, AK1, AK2}. The first case admits the usual (bright) soliton solutions and the second has self-modulated solitons that are solitonic envelops propagating on a background.

The NLS model on the half-line has also been extensively studied in the literature. 
Sklyanin was the first to derive integrable BCs of the model on an interval. His work was then followed by important contributions \cite{fokas1989initial, Tarasov, HH1, fokas1996linearization} in which the main idea was to incorporate the integrable BCs into the powerful IST framework.
The unified transform method has also been applied to the half-line NLS model \cite{fokas2002integrable, fokas2005nonlinear}. In particular, the authors had the Robin BCs as linearizable BCs, and derived asymptotic soliton solutions at large time. 
In spite of a good understanding of the model, exact soliton solutions of NLS on the half-line were only obtained rather recently in \cite{biondini2009solitons} where a nonlinear mirror-image technique was applied. The method consists in extending the half-line space domain to the whole axis. This reflects the space-inverse symmetry of NLS and allows to obtain solutions of the model using the usual IST by uniquely looking at the positive semi axis. The technique was successfully applied to the vector NLS model \cite{CZ},  and was used to obtain boundary-bound solitons \cite{bion2} that are static solitons subject to the BCs. Note that a recent study of the model was reported in \cite{DP1} following a functional analytical approach. 

Our approach to solving the NLS equation on the half-line is based on the theory of integrable BCs. It is well-known that soliton solutions of integrable PDEs can be obtained in an algebraic fashion using the Darboux-dressing transformations (DTs), cf.~\cite{ZS4, ZS5, matveev1991darboux}.  The essential idea of our method is to ``dress" the integrable boundary structures at the boundary point $x=0$ in preserving the integrability. In other words, we are dressing the system at the boundary.  Two classes of seed solutions are considered: the zero seed solution, which correspond to the usual (bright) solitons, and the non-zero seed solution, which has self-modulated solitons on a constant background.  In both cases, soliton solutions on the half-line, as well as boundary-bound solitons,  arise directly in the boundary-dressing process. One of the particular advantage of our method is that it does not require  extension of the space domain to the whole axis. It also admits a natural IST interpretation: the ``boundary profile" at $x=0$ is encoded into scattering data that evolve linearly on the semi axis, then the half-line solutions can be obtained in solving the inverse problems.  Although the main content of the paper is focusing on a particular model that is the NLS equation on the half-line, the scope of the paper is to address to a rather general question of how to determine exact solutions of integrable PDEs accompanied with integrable BCs.

The paper is organized as follows: we review Sklyanin's approach to integrable BCs and the DT approach to generating soliton solutions in Section $2$ and $3$ respectively. This lays the basis for the boundary-dressing method presented in Section $4$. In Section $5$ and $6$, the method is applied to two classes of soliton solutions of NLS. An IST interpretation of  boundary-dressing method the is given in Section $7$. 

\section{Sklyanin's approach to integrable BCs for NLS}
\label{sec:sf1}
We start by a brief summary of Sklyanin's formalism \cite{sklyanin1987boundary} (see also \cite{ACC} for recent studies) to derive integrable BCs for the NLS equation on an interval. This formalism is based on the Hamiltonian formulation of integrable PDEs, cf.~\cite{faddeev}. 
\subsection{Classical  $r$-matrix structure}
The NLS equation is the result of the compatibility between the linear differential equations
\begin{equation}
  \label{eq:laxp}
U\, \phi=\phi_x\,, \quad
   V\, \phi=\phi_t\,. 
\end{equation} 
Here, the matrix-valued functions $U$ and $V$, known as the Lax pair, depend on $x, t$ and also on a spectral parameter $
\lambda$\footnote{We drop the the $x$, $t$ and $\lambda$ dependence for conciseness unless there is ambiguity.}. They are in the forms
\begin{equation}
  U = -i \lambda \sigma_3 +Q \,,\quad V = -2i\lambda^2\sigma_3+2\lambda \, Q -iQ_x\,\sigma_3-i Q^2\,\sigma_3\,,
\end{equation}
where
\begin{equation}
  \label{eq:laxp2}
  \sigma_3=\bma 1 & 0\\ 0 & - 1\ema \,, \quad Q=\bma 0 & q \\ -\bar{q} & 0 \ema\,.\quad   
\end{equation}
We will call $U$ and $V$ the  $x$-part and $t$-part of the Lax pair respectively. 

Now, restricting the space domain to an interval $[x_-,x_+]$ with $x_\pm$ being the two ends. A particularly important quantity called  {\it monodromy matrix} $T(\lambda)$, obtained from the $x$-part of the Lax pair at a fixed time $t_0$,  is the  following ordered exponential function  
\begin{equation}
  T(\lambda) =  \overset{\curvearrowleft}{\exp}  \int^{x_+}_{x_-} U(\xi,t_0,\lambda)\, d\xi \, ,
\end{equation}
for it encodes  the spectral properties in  the process of the direct scattering in the IST. 
In the  Hamiltonian formulation of  the NLS equation \cite{faddeev}, 
two monodromy matrices satisfy
\begin{equation}
  \label{eq:sk1}
  \{T(\lambda) \underset{'}{\otimes}    T(\eta)  \} = [r(\lambda -\eta),   T(\lambda)  \otimes  T(\eta)]\,,
\end{equation}
where $\otimes$ is the standard tensor product for matrices, and the operation $\{\cdot\, \underset{'}{\otimes}\, \cdot\}$ stands for the Poisson bracket in tensor-product form. 
The quantity $r(\lambda)$ is called the classical $r$ matrix, and it satisfies the classical Yang-Baxter equation
\begin{equation}
  \label{eq:cyb}
    [r_{12}(\lambda - \eta) , r_{13}(\lambda) + r_{23}(\eta)]+ [r_{13}(\lambda) , r_{23}(\eta)] = 0\,.
\end{equation}
It is in the form\footnote{The explicit form of $r(\lambda)$ depends on the Poisson structure of the model. In this paper, we use
\[
    \{q(x), \bar{q}(y)\} = -i \delta(x-y)\,. 
\]

 }
\begin{equation}
  r(\lambda) = \frac{1}{2\lambda}\, \mathcal{P} \,, \quad \mathcal{P} =\bma 1 & 0 & 0 & 0\\ 0 & 0 & 1 & 0\\0 & 1 & 0 & 0\\0 & 0 & 0 & 1 \ema \,,
\end{equation}
where $\mathcal{P}$ is the permutation matrix acting on $\CC^2\otimes \CC^2$. 
The formula \eqref{eq:sk1} is a universal integrable structure characterizing wide classes of classical integrable models \cite{faddeev}. 
It is entirely determined by the form of $U$, and is independent of  conditions imposed at the boundaries $x_\pm$. It can be used to prove the integrability of the NLS equation 
under the quasi-periodic BCs~\cite{faddeev}.  However, for generic BCs, the above relations are not enough to guarantee the  integrability. Another equation involving the BCs is needed.  
\begin{rmk}
  The  quasi-periodic BCs for NLS read
  \begin{equation}
     q(x_-)=e^{i\varphi}q(x_+)\,,
  \end{equation}
where $0\leq \varphi <2 \pi$. Both the vanishing and  non-vanishing BCs can be seen as the special cases of   the  quasi-periodic BCs \cite{faddeev}.
\end{rmk}
\subsection{Classical reflection equation}
We proceed to the Sklyanin's formalism. Recall that  $x_-$ and $x_+$ are the boundaries. One can  show, if there exist nonsingular spectrum-dependent matrices $K_\pm(\lambda)$ satisfying 
\begin{align}
\displaystyle  \label{eq:cre}
  \lbrack r(\lambda-\eta)\, , \,  &  K_\pm (\lambda)  \otimes  K_\pm(\eta) \rbrack =  \left(  I  \otimes K_\pm(\eta)\right)  \, r(\lambda+\eta)\,   \left(K_\pm(\lambda)  \otimes  I\right)\nonumber \\
&- \left(K_\pm(\lambda)  \otimes  I\right)\, r(\lambda+\eta) \, \left(  I  \otimes K_\pm(\eta)\right)   \,,
\end{align}
with $I$ being the identity matrix, then the integrable algebra \eqref{eq:sk1} is deformed to 
\begin{align}
  \label{eq:cra}
  \{\mathcal{T} (\lambda) \underset{'}{\otimes}  \mathcal{T}   (\eta)  \} = &[r(\lambda -\eta) \,, \,\mathcal{T}(\lambda)  \otimes  \mathcal{T}(\eta)] + (\mathcal{T} (\lambda) \otimes I) \, r(\lambda+\eta) \, (I \otimes \mathcal{T} (\eta) ) \,   \nonumber \\
&-  (I \otimes \mathcal{T} (\eta) ) \,r(\lambda+\eta)  \,(\mathcal{T} (\lambda) \otimes I)\,,
\end{align}
with $
\mathcal{T}(\lambda) =  T(\lambda) \,  K_-(\lambda) \, T^{-1}(-\lambda)
$.
In this setting, a generating function $\tau(\lambda)$  can be constructed as
\begin{equation}
  \label{eq:gf}
  \tau(\lambda) = \tr \left( K_+(\lambda )\, \mathcal{T}(\lambda) \right) \,,
\end{equation}
and   satisfies
\begin{equation}
  \{\tau(\lambda)\,,\,\tau(\eta)\}=0\,.
\end{equation}
 The function $\tau (\lambda)$ is interpreted as a generating function of commuting integrals of motion. In order that $\tau(\lambda)$ generates infinitely many conserved quantities, it should be time-independent, which is equivalent to the following conditions 
 \begin{align}
\label{eq:NV1}
  K_+(\lambda) \, V(\lambda, x_+) =V(-\lambda, x_+)  \, K_+(\lambda)\,,\\
\label{eq:NV2}
  K_-(\lambda) \, V(-\lambda, x_-) =V(\lambda, x_+)  \, K_-(\lambda)\,. 
\end{align}
Therefore, the complete integrability of the NLS equation in the presence of boundaries is obtained. The integrable BCs are encoded into the  boundary matrices $K_\pm$ that satisfy both Eq~\eqref{eq:cre} and Eqs~(\ref{eq:NV1}, \ref{eq:NV2}).
 
\begin{rmk}
\label{rm:21} It is worth noting that  as $x_\pm$ tend to $ \pm \infty$,  both vanishing BCs and non-vanishing BCs  have   $K_\pm \propto $ some constant  diagonal matrices as solutions. They are trivial solutions of Eqs~(\ref{eq:cre}) and (\ref{eq:NV1}, \ref{eq:NV2}). Both cases are thus integrable. 
\end{rmk}

\section{DTs and soliton solutions }
DT approach is a direct method for generating  soliton solutions  of integrable PDEs. The underlying structures are connected to the IST involving studies of the scattering properties of the Lax system \eqref{eq:laxp}.  Here we  are mainly following  the notations used in  \cite{matveev1991darboux}. 
\subsection{One-fold DTs}
DTs are  gauge transformations preserving the forms of the Lax pair. For integrable PDEs, B\"acklund transformations are results of DTs at the level of solutions, as an ``old"  solution is transformed into a ``new"  solution. For notational purpose, we  use $U[0]$, $V[0]$ and $\phi[0]$ to denote the {\em undressed} Lax system~\eqref{eq:laxp}.  

DT for NLS can be defined as follows: suppose there is a gauge-like transformation  
 \begin{equation}
\label{eq:gt1}
   \phi[1] = D[1]\, \phi[0]\,. 
 \end{equation}
Now $\phi[1]$ satisfies the newly transformed system
\begin{equation}
\label{eq:gt2}
  U[1]\,\phi[1] =\phi[1] _x\,,\quad V[1]\,\phi[1] =\phi[1] _t\,.
\end{equation}
The structure of $U[1], V[1]$ is required to be identical to that of $U[0], V[0]$. 
They are connected by
\begin{equation}
\label{eq:gt3}
  D[1]_x =  D[1]\,U[0]-U[1]\,D[1]\,, \quad  D[1]_t = D[1]\,V[0]-V[1]\,D[1]\,.
\end{equation}
The matrix $D[1]$ is called dressing matrix. 
There are several equivalent representations  of  $D[1]$, 
 and here we adopt the polynomial (in $\lambda$) form  
\begin{equation}
  \label{eq:dar1}
 D[1] =\left(\lambda-\bar{\lambda}_1\right) + \left( \bar{\lambda}_1-\lambda_1\right) P[1]\,,
\quad P[1] = \frac{\psi_1 \psi_1^\dagger}{\psi_1^\dagger\psi_1}\,.\end{equation}
The $2$-vector $\psi_1 = (\mu_1,\nu_1)^\intercal$ is a special solution of the undressed Lax system~\eqref{eq:laxp}  at $\lambda = \lambda_1$, $\psi_1^\dagger$ denotes the transpose conjugate of $\psi_1$.
 The dressing matrix $D[1]$ defines  a one-fold Darboux transformation, as it adds a pair of zeros to the undressed Lax system\footnote{This can be easily seen by taking the determinant of $D[1]$:
 \begin{equation}
   \det D[1] = (\lambda -\lambda_1)\,(\lambda - \bar{\lambda}_1)\,.
 \end{equation}}.

Putting  $D[1]$ into Eq~(\ref{eq:gt3}) gives the form of  $Q[1]$ in terms of $Q[0]$ and $P[1]$:
\begin{equation}
  \label{eq:sol1}
  Q[1] = Q[0] - i(\lambda_1 - \bar{\lambda}_1) [\sigma_3, P[1]]\,.
\end{equation}
This is the  {\em reconstruction formula}, and a one-soliton solution $q[1]$ can be easily obtained. 
The function $q[0]$ is commonly known as {\em seed solution}. In the process of DTs, the choice of $q[0]$ is important, because the Lax system~\eqref{eq:laxp} cannot be integrated with  arbitrary $q[0]$. In practice, we choose the zero seed solution $q[0]=0$ and the non-zero constant background seed solution $q[0] =  \rho \, e^{2i\rho^2 t}    $, $\rho>0$. Both cases will be studied later in the presence of integrable BCs.

\subsection{$N$-soliton solutions}
$N$-fold DTs can be constructed by iteration. 
Assume that there exist $N$ linearly-independent special solutions $\psi_{j} = (\mu_j,\nu_j)^\intercal$ of the undressed Lax system~\eqref{eq:laxp} evaluated at  $\lambda_j$, $j = 1\,\dots\,N$,  then the $N$-fold dressing matrix $D[N]$ is in the form
\begin{equation}
\label{eq:dbnf2}
  D[N] = \left(( \lambda-\bar{\lambda}_N) + ( \bar{\lambda}_N-\lambda_N)\, P[N] \right) \, \cdots \left(( \lambda-\bar{\lambda}_1) \right. +   \left.( \bar{\lambda}_1-\lambda_1)\, P[1] \right)\,, 
\end{equation}
where
\begin{equation}
\label{eq:prj11}
  P[j] = \frac{\psi_j [j-1] \, \psi^\dagger_j [j-1]}{\psi^\dagger_j [j-1] \, \psi_j [j-1]}\,, \quad \psi_j [j-1]  = D[j-1]\big\rvert_{\lambda=\lambda_j}\,  \psi_{j}\,.
\end{equation}
A series expansion of      $D[N] $ in $\lambda$ leads to
\begin{equation}
\label{eq:dbnf1}
  D[N] = \lambda^N +\lambda^{N-1 }\,\Sigma_1 + \lambda^{N-2 }\,\Sigma_2 \cdots+ \Sigma_N\,,
\end{equation}
with the matrices $\Sigma_j$, $j=1\,\dots\,N$ to be determined. 
Inserting $\phi[N]=D[N] \, \phi[0]$ into
\begin{equation}
  U[N]\,\phi[N]\, = \phi[N]_x\,,\quad V[N]\,\phi[N]\, = \phi[N]_t\,,
\end{equation}
and taking  account of the forms of $Q[0]$ and $Q[N]$, one can obtain the reconstruction formula for $N$-soliton solutions 
\begin{equation}
\label{eq:solqn}
q[N]=q[0]+2 i\,  \Sigma_1^{(1,2)} \,,
\end{equation}
where $\Sigma_1^{(1,2)}$ is the $(1,2)$ entry of  $\Sigma_1$. It can be put into the following compact form
\begin{equation}
\label{eq:solqn}
q[N]=q[0]+2i \frac{\Delta_1}{\Delta_2} \,,
\end{equation}
where\begin{equation}
\Delta_1=\begin{vmatrix}
  -\lambda_{1}^{N}\mu_1 &  \cdots & -\lambda_{N}^{N}\mu_N  &  \bar{\lambda}_{1}^{N}\bar{\nu}_1  & \cdots & \bar{\lambda}_{N}^{N}\bar{\nu}_N\\
           \lambda_{1}^{N-2}\nu_1 &  \cdots & \lambda_{N}^{N-2}\nu_N  &  \bar{\lambda}_{1}^{N-2}\bar{\mu}_1  & \cdots & \bar{\lambda}_{N}^{N-2}\bar{\mu}_N\\
                            \cdots & \cdots & \cdots & \cdots &  \cdots & \cdots\\
                            \nu_1 &  \cdots & \nu_N  &  \bar{\mu}_1  & \cdots & \bar{\mu}_N \\
     \lambda_{1}^{N-1}\mu_1 &  \cdots & \lambda_{N}^{N-1}\mu_N  &  -\bar{\lambda}_{1}^{N-1}\bar{\nu}_1  & \cdots & -\bar{\lambda}_{N}^{N-1}\bar{\nu}_N\\
                            \cdots & \cdots & \cdots & \cdots &  \cdots & \cdots\\
                            \mu_1 &  \cdots & \mu_N  &  -\bar{\nu}_1  & \cdots & -\bar{\nu}_N                  \end{vmatrix}\,,
\end{equation}
\begin{equation}
\label{eq:delta2}
\Delta_2=\begin{vmatrix}
                            \lambda_{1}^{N-1}\nu_1 &  \cdots & \lambda_{N}^{N-1}\nu_N  &  \bar{\lambda}_{1}^{N-1}\bar{\mu}_1  & \cdots & \bar{\lambda}_{N}^{N-1}\bar{\mu}_N\\
                            \cdots & \cdots & \cdots & \cdots &  \cdots & \cdots\\
                            \nu_1 &  \cdots & \nu_N  &  \bar{\mu}_1  & \cdots & \bar{\mu}_N\\
                            \lambda_{1}^{N-1}\mu_1 &  \cdots & \lambda_{N}^{N-1}\mu_N  &  -\bar{\lambda}_{1}^{N-1}\bar{\nu}_1  & \cdots & -\bar{\lambda}_{N}^{N-1}\bar{\nu}_N\\
                            \cdots & \cdots & \cdots & \cdots &  \cdots & \cdots\\
                            \mu_1 &  \cdots & \mu_N  &  -\bar{\nu}_1  & \cdots & -\bar{\nu}_N
          \end{vmatrix} \,.
        \end{equation}
This  formula will be used later in the computation of half-line solutions of the NLS equation.  A brief proof  is  given in  Appendix~\ref{ap:1}.

 \section{Dressing integrable BCs}
\label{sec:DBBC}
Now we proceed to the main results of the paper. The Sklyanin's formalism, described in Section~\ref{sec:sf1}, characterizes the integrability of the NLS equation on an interval $x_- \leq x \leq x_+$. An half-line problem can be easily realized as a special case of the interval problem  by setting $x_{-}= 0$ and  $x_+\to \infty $.

Our aim is to compute exact solutions of the NLS equation on the half-line without special treatments such as extending the space domain into the whole line, or performing asymptotic analysis at large $t$. One knows that the integrable BCs are completely determined by the $t$-part of the Lax pair through Eqs~(\ref{eq:NV1}, \ref{eq:NV2}) (with the boundary matrices also satisfying Eq~\eqref{eq:cre}), and that the DT approach to generating soliton solutions can, in principle, be defined in any space-time domain. The problem  remains to find suitable DT for the NLS model on the half-line, with the $t$-part of the Lax pair  satisfying the constraints~(\ref{eq:NV1}, \ref{eq:NV2}).


\subsection{Integrable BCs and  boundary matrices}
According to the Sklyanin's formalism, the boundary matrices $K_\pm(\lambda)$ 
can be  treated  {\em separately}. In the rest of the paper, we assume that   $K_+ $ is proportional to some constant diagonal matrices, which are trivial solutions of Eqs~\eqref{eq:NV2} and \eqref{eq:cre}. This assumption can be justified since we only consider two classes of solutions: solutions under the vanishing BCs as $x\to \infty$
\begin{equation}
  \lim_{x\to\infty} q = 0\,,  \quad \lim_{x\to\infty} q_x = 0\,,
\end{equation}
which have  $K_+ \propto I$,  and solutions under  the ``constant" background condition (non-vanishing BCs)
\begin{equation}
  \lim_{x\to\infty} q = \rho \,e^{2 i \rho^2 t}, \quad \lim_{x\to\infty} q_x=0\,,
\end{equation}
with $\rho $ being a positive real constant acting as the background, which have $K_+ \propto \sigma_3$.  

For conciseness, we use  $K$  instead of  $K_-$, then one has to solve the boundary constraint
\begin{equation}
\label{eq:dbm1}
  K(\lambda)\, V(-\lambda)\big\rvert_{x=0} =   V(\lambda)\big\rvert_{x=0}\,K(\lambda)\,,
\end{equation}
with $K(\lambda)$ also satisfying~\eqref{eq:cre}.   This constraint corresponds to the {\em  linearizable BCs} in the  Fokas' unified transform method\cite{fokas2002integrable, fokas2005nonlinear}.  
A class of solutions is the  Robin BCs:
\begin{equation}
\label{eq:rtbcs1}
 q_x\big\rvert_{x=0}= \alpha \,q\big\rvert_{x=0}\,, \quad \alpha \in \RR\,,  
\end{equation} 
which has   
the boundary matrix $K(\lambda)$    in the form
\begin{equation}
\label{eq:formkf}
  K(\lambda )  = \bma f_\alpha(\lambda)  & 0 \\ 0 & 1 \ema \,,\quad f_\alpha(\lambda) = \frac{i\,\alpha+2\lambda }{i\, \alpha-2\lambda }\,.
\end{equation}
One can easily check that $K^{-1}(\lambda) = K(-\lambda)$. The real parameter $\alpha$ controls the boundary behavior. The limiting cases of $\alpha$ give rise either to  Dirichlet BCs
\begin{equation}
  q\big\rvert_{x=0} = 0\,, \quad \alpha \to \infty\,,\quad K =  I\,,
\end{equation}
or to the Neumann BCs
\begin{equation}
  q_x\big\rvert_{x=0} = 0\,, \quad \alpha = 0 \,,\quad K = -\sigma_3\,. 
\end{equation}
We exclude the special cases $\lambda = \pm \frac{ i \alpha}{2}$ due to the non-degeneracy of $K(\lambda)$.
One can check that 
the boundary matrix $K(\lambda)$ defined in~\eqref{eq:formkf}  satisfies the classical reflection equation~\eqref{eq:cre}, which concludes that the Robin BCs \eqref{eq:rtbcs1} are integrable BCs, cf.~\cite{sklyanin1987boundary, BT1, Tarasov}. 
\begin{rmk}
It is interesting to see that the Robin BCs are not the  only solutions of the boundary constraint \eqref{eq:dbm1}. For instance, one can have
\begin{equation}
  \label{eq:obcs1}
    q\big\rvert_{x=0}= \rho\, e^{i\theta} \,, \quad  \bar{q}_x\big\rvert_{x=0} = e^{-2i\theta} \, q_x\big\rvert_{x=0}\,, 
  \end{equation}
  as the BCs, with $\rho \neq 0$, $\theta \in\RR $. The corresponding boundary matrix $K(\lambda)$  is in the form  
  \begin{equation}
\label{eq:obcs2}    K(\lambda) = \bma  1 & 2ie^{i\theta}\frac{\lambda\rho}{  \rho^2-2\lambda^2} \\2ie^{-i\theta}\frac{\lambda\rho}{  \rho^2-2\lambda^2} & 1 \ema\,. 
  \end{equation}
  This solution fits perfectly into the definition of linearizable BCs in the unified transform method. They can be {\em in theory} used in the simultaneous treatments of both the $x$-part and $t$-part of the Lax pair, in which the boundary matrix $K(\lambda)$ will imply certain reductions of the scattering coefficients of the initial-boundary data. However,  the matrix $K(\lambda)$ in this case does not satisfy the classical reflection equation \eqref{eq:cre}. Therefore, the BCs \eqref{eq:obcs1}  are not  integrable.
\end{rmk}

\subsection{Integrable BCs as Darboux-B\"acklund transformations}
The boundary constraint \eqref{eq:dbm1}  with the boundary matrix $K(\lambda)$ defined in \eqref{eq:formkf} can be interpreted as a DT for certain Lax pair with its space domain extended to the whole axis.

Imposing the following relation
\begin{equation}
  \label{eq:xis}
  \widetilde{Q}(x,t) = Q(-x,t)  \,.
\end{equation}
One can  verify that
\begin{equation}
  \widetilde{V}\big\rvert_{x=0} = \sigma_3\, V(-\lambda)\big\rvert_{x=0}\,\sigma_3\,,
\end{equation}
with $\widetilde{V}$ defined in terms of $\widetilde{Q}(x,t)$. The relation \eqref{eq:xis} is actually a B\"acklund transformation of NLS due to its space-inverse symmetry. Taking account of \eqref{eq:dbm1}, 
the above formula can be expressed as a DT for $V$ at $x=0$:
\begin{equation}
  \label{eq:xis1}
  D(\lambda)\, V(\lambda)\big\rvert_{x=0} - \widetilde{V}(\lambda)\big\rvert_{x=0}\,D(\lambda) = 0\, ,\quad  D(\lambda) = \sigma_3 \,K(-\lambda)\,.
\end{equation}
The boundary constraint \eqref{eq:dbm1} is just reformulated as a DT \eqref{eq:xis1}, with $Q $ transformed to $\widetilde{Q}$ as the  action of the DT.
In order that the DT \eqref{eq:xis1} is also defined for the $x$-part of the Lax pair $U$,  one needs to introduce a  functional $\Pi$ in the form
\begin{equation}
  \label{eq:epwl1}
  \Pi(x,t,\lambda) = {\mathcal H}(x)\,Q(x,t)+ {\mathcal H}(-x)\,D(-\lambda)\,\widetilde{Q}(x,t) \,D(\lambda)\,,
\end{equation}
where ${\mathcal H} $ is the Heaviside function. Now the functional $\Pi$, combining both $Q$ and $\widetilde{Q}$, is a field defined on the whole real axis. It satisfies
\begin{equation}
  \widetilde{\Pi}(x,t,\lambda) := \Pi(-x,t,-\lambda) = D(\lambda)\, \Pi(x,t,\lambda)\, D(-\lambda)\,.
\end{equation}
Using the extended field $\Pi$ instead of $Q$ in the Lax pair, \ie
\begin{equation}
  \label{eq:laxpet}
  U= -i\lambda\,\sigma_3+\Pi\,,\quad V = -2i\lambda^2\,\sigma_3+2\lambda \, \Pi -i\Pi_x\,\sigma_3-i \Pi^2\,\sigma_3\,, 
\end{equation}
one can  verify the following relations 
\begin{equation}
  D(\lambda)\, U(\lambda)\big\rvert_{x=0}- \widetilde{U}(\lambda)\big\rvert_{x=0}\,D(\lambda) = 0\, ,  \quad
  D(\lambda)\, V(\lambda)\big\rvert_{x=0} - \widetilde{V}(\lambda)\big\rvert_{x=0}\,D(\lambda) = 0\,  . 
\end{equation}
These relations incorporate the boundary constraint \eqref{eq:dbm1} into the Lax pair $U, V$ with an extended field $\Pi$. The equivalence between the extended Lax pair   \eqref{eq:laxpet} and the integrable BCs is thus established. The apparent advantage of the introduction of the extended field $\Pi$ is that it allows us to perform the usual IST. This corresponds to the mirror-image method for solving the NLS equations \cite{biondini2009solitons}.   
However, there are some analytic difficulties in dealing with the functional field $\Pi$ as it contains the Heaviside function which requires special care  at $x=0$. Moreover, extending the space domain to the whole real axis does not seem to be the most natural approach to a half-line problem. 
\subsection{Dressing the boundary}
The construction of exact solutions of the NLS equation on the half-line is based on the following steps. At the seed solution level, taking account of the boundary constraint \eqref{eq:dbm1},  given
$  \psi_1(\lambda_1) $ as a special solution of the undressed Lax system $U[0]$ and $V[0]$,
if there exist another {\em paired} special solution $ \psi_2(\lambda_2) $ obeying
\begin{equation}
  \label{eq:431}
  \psi_2(0,t,\lambda_2) = K(\lambda_2) \, \psi_1(0,t,\lambda_1)\,,\quad \lambda_2 = -\lambda_1\,,\quad \bar{\lambda}_1 \neq -\lambda_1 \,,
\end{equation}
then the seed solution $q[0]$ satisfies the BCs  imposed by \eqref{eq:dbm1}. 
This can be easily understood by looking at the $t$-part equation of the Lax system at $x=0$
\begin{equation}
  V[0](0,t,\lambda_j)\,\psi_j(0,t,\lambda_j) =\psi_{j\,t}(0,t,\lambda_j)\,,\quad  j=1,2\,.
\end{equation}
By inserting the relation \eqref{eq:431} into the previous equations one gets
\begin{equation}
K(\lambda_j)\,  V[0](-\lambda_j)\,K(-\lambda_j) \,\psi_j = V[0](\lambda_j)\,\psi_j = \psi_{j\,t}(\lambda_j)\,, \quad  j =1,2\,,
\end{equation}
at $x=0$, which implies  $V[0]$ is subject to the boundary constraint \eqref{eq:dbm1}. Note that we exclude for the moment the case where $\lambda_1$ is a pure imaginary (because of the condition $\bar{\lambda}_1 \neq -\lambda_1$). 
In fact, the existence of \eqref{eq:431} is a strong condition which allows to construct soliton solutions satisfying   the BCs  imposed by \eqref{eq:dbm1}. 
\begin{proposition}[Dressing the boundary]
  \label{prop:1}
Consider the half-line NLS model with Robin BCs  \eqref{eq:rtbcs1}. Assume that there exist paired special solutions $\{\psi_1,\widehat{\psi}_1\}$ of  the undressed Lax system~\eqref{eq:laxp} associated with the parameters $\{\lambda_1, \widehat{\lambda}_1\}$  such that 
  \begin{equation}
\label{eq:cond41}
    \widehat{\psi}_1(0,t,\widehat{\lambda}_1) = K(\widehat{\lambda}_1)\,\psi_1(0,t,\lambda_1)\,,\quad \widehat{\lambda}_1 = -\lambda_1\,,\quad \bar{\lambda}_1  \neq -\lambda_1\,,
  \end{equation}
where $K(\lambda)$ is  given in \eqref{eq:formkf},  then a two-fold DT using such pair  leads to a $V[2]$ satisfying  
\begin{equation}\label{eq:v2v2}
K(\lambda)\,V[2](-\lambda)\rvert_{x=0} =  V[2](\lambda)\,K(\lambda)\rvert_{x=0}\,. 
\end{equation}
The so-constructed  $q[2]$ satisfies the Robin BCs. We use $\widehat{q}[1]$ to denote such $q[2]$. 
\end{proposition}
\prf In order that  $\widehat{q}[1]$  is an exact solution of the NLS equation on the half-line, we need to prove the relation \eqref{eq:v2v2}. Let $D[2](\lambda)$ be the dressing matrix constructed using  $\{\psi_1,\widehat{\psi}_1\}$. One knows that $V[2](\lambda)$ and $V[0](\lambda)$ are connected by
\begin{equation}
  V[2](\lambda)=D[2]_t(\lambda)\, D[2]^{-1}(\lambda)+D[2](\lambda)\,V[0](\lambda)\,D[2]^{-1}(\lambda)\,.
\end{equation}
One can easily verify that if $D[2](\lambda)$ satisfies
\begin{equation}
  \label{eq:432}
  D[2](\lambda)\,K(\lambda)\rvert_{x=0} = K(\lambda)\,D[2](-\lambda)\rvert_{x=0}\,,
\end{equation}
then $V[2](\lambda)$ satisfies  \eqref{eq:v2v2}. Multiplying both sides of Eq~\eqref{eq:432} by an irrelevant factor $(i\,\alpha-2\lambda)$. Since $D[2]$ can be expressed as a matrix  polynomial of degree $2$ in $\lambda$ and that   $(i\alpha-2\lambda) K(\lambda) = i\, \alpha I+2\lambda\sigma_3 $, the {\it l.h.s.} and {\it r.h.s.} of \eqref{eq:432} are thus matrix  polynomials of degree $3$. We use $L(\lambda)$ and $R(\lambda)$ to denote them
\begin{align}
  L(\lambda)=&D[2](\lambda)\,K(\lambda) = \lambda^3L_0+\lambda^2L_1+\lambda^2L_2+L_3\,,\\
    R(\lambda)=&K(\lambda)D[2](-\lambda)= \lambda^3R_0+\lambda^2R_1+\lambda^2R_2+R_3\,. 
\end{align}
Clearly, $L_0=R_0$, $L_3=R_3$, and $L_1,L_2,R_1,R_2$ can be determined by the zeros of $L(\lambda), R(\lambda)$ and the associated kernel vectors, cf.~\cite[Chapiter (3.10)]{babelon}.  Following the zeros of  $D[2](\lambda)$, \ie $D[2](\lambda_1)\psi_1=0$ and $ D[2](\widehat{\lambda}_1)\widehat{\psi}_1=0$, and the relation between $\psi_1$ and $\widehat{\psi}_1$ \eqref{eq:cond41}, one has
\begin{align}
R(\lambda)\rvert_{\lambda=-\lambda_1}\psi_1= 0\,,\quad   L(\lambda)\rvert_{\lambda=-\lambda_1}\psi_1= 0\,,\quad R(\lambda)\rvert_{\lambda={\lambda}_1}\widehat{\psi}_1= 0\,,\quad   L(\lambda)\rvert_{\lambda={\lambda}_1}\widehat{\psi}_1= 0\,,
\end{align}
evaluated at  $x=0$. 
Moreover, let $\varphi_1 = \sigma_2 \bar{\psi}_1$, $\sigma_2= \bma 0 & -i \\ i & 0 \ema$, one can verify (see also Appendix \ref{ap:1}) 
\begin{align}
R(\lambda)\rvert_{\lambda=-\bar{\lambda}_1}\varphi_1= 0\,,\quad   L(\lambda)\rvert_{\lambda=-\bar{\lambda}_1}\varphi_1= 0\,,\quad R(\lambda)\rvert_{\lambda=\bar{\lambda}_1}\widehat{\varphi}_1= 0\,,\quad   L(\lambda)\rvert_{\lambda=\bar{\lambda}_1}\widehat{\varphi}_1= 0\,, 
\end{align}
at  $x=0$, where $\widehat{\varphi}_1$ is defined in the same way as $\varphi_1$. The above formulae show that  $L(\lambda)$ and $R(\lambda)$ share the same zeros and the associated kernel vectors, which, in turn, implies that   $L(\lambda)=R(\lambda)$. This completes the proof. 
\finprf

The above construction is the realization of ``dressing the boundary". In fact, the existence of the paired special solutions $\{\psi_1(\lambda_1),\widehat{\psi}_1(\widehat{\lambda}_1)\}$  implies that the  $t$-part of undressed Lax pair $V[0]$ satisfies the boundary constraint \eqref{eq:dbm1}. Thus the seed solution $q[0]$ is ``presumed" to be subject to the integrable BCs \eqref{eq:rtbcs1}. Dressing  $V[0]$ using the pair  $\{\psi_1(\lambda_1),\widehat{\psi}_1(\widehat{\lambda}_1)\}$   preserves  the boundary constraint \eqref{eq:dbm1}, hence the BCs \eqref{eq:rtbcs1}. The so-constructed solution $\widehat{q}[1]$ represents a one-soliton solution on the half-line, although two special solutions are involved. This can be naturally understood as follows:  the paired special solutions $\{\psi_1(\lambda_1),\widehat{\psi}_1(\widehat{\lambda}_j)\}$ create asymptotically at large  $t$ two independent solitons with opposite vilocities, as time evolves from $-\infty$ to $\infty$, only one soliton remains in $x\geq 0$, and  the BCs correspond to the interaction of the two solitons.  Similar phenomena happen both in the mirror-image approach to computing exact half-line solutions \cite{biondini2009solitons, CZ}, and in the unified transform method when performing the large-time analysis \cite{fokas2002integrable, fokas2005nonlinear}. Note that our construction of exact solutions can be restricted to the half-line, as opposed to the mirror-image technique where an extended functional to the whole axis is needed. It also turns out our construction possesses a natural IST interpretation (see Section $7$).

One can repeat dressing the boundary using $N$ paired special solutions.
\begin{proposition}[$N$-soliton solutions]
\label{prop:NSS}Consider the half-line NLS model with Robin BCs \eqref{eq:rtbcs1}. Assume that there exist $N$ paired special solutions $\{\psi_j(\lambda_j),\widehat{\psi}_j(\widehat{\lambda}_j)\}$, $j=1,\dots, N$, of  the undressed Lax system~\eqref{eq:laxp} such that
  \begin{equation}
\label{eq:cond4111}
    \widehat{\psi}_j(0,t,\widehat{\lambda}_j) = K(\widehat{\lambda}_j)\,\psi_j(0,t,\lambda_j)\,,\quad \widehat{\lambda}_j = -\lambda_j\,,\quad \bar{\lambda}_j  \neq -\lambda_j\,,\quad \widehat{\lambda}_k\neq \lambda_j\,,
  \end{equation}
where $K(\lambda)$ is given \eqref{eq:formkf}.  Then, the so-constructed  $q[2N]$  corresponds to an $N$-soliton solution of NLS on the half-line. We use $\widehat{q}[N]$ to denote such $q[2N]$. 
\end{proposition}
The requirement $\widehat{\lambda}_k\neq \lambda_j$ ensures that all the special solutions are independent. By construction, the integrable structures such as the boundary constraint \eqref{eq:dbm1} and the BCs \eqref{eq:rtbcs1} are preserved at each step of the dressing. It remains to find such pairs  $\{\psi_j(\lambda_j),\widehat{\psi}_j(\widehat{\lambda}_j)\}$. 
Also note that the case where $\psi_j$ has pure imaginary spectral parameter is not included into the above formalism. These problems are considered in the next two Sections when we deal with the concrete examples. 
\section{Soliton solutions of NLS on the half-line}
\label{sec:nls1}
We consider  the zero seed solution  $  q[0] = 0$, which has the vanishing BCs as $x\to \infty$. The special case where $\psi_j$ has pure imaginary parameter is also treated. The latter corresponds to static solitons bounded to the boundary. 
\subsection{Soliton solutions}
\begin{figure}[h!]
  \centering
  \includegraphics[width=0.4\linewidth]{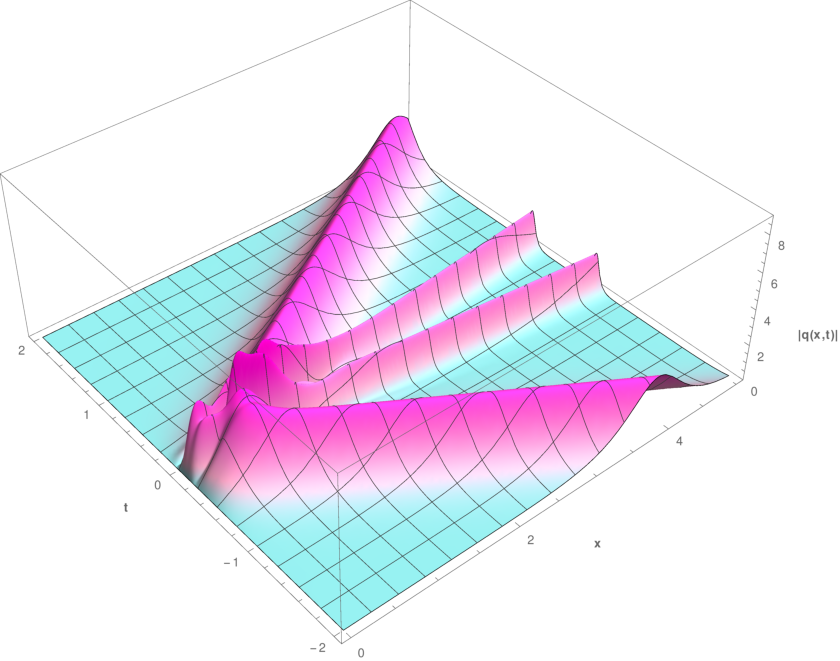} \hspace{.4cm}
  \includegraphics[width=0.4\linewidth]{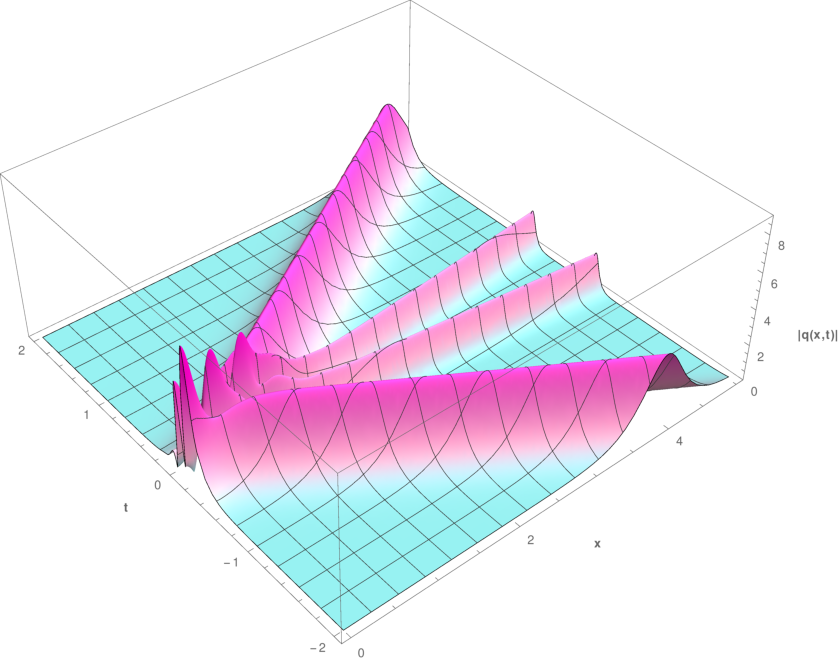}
  \caption{Two-soliton solution satisfying the Dirichlet BCs (left) and Neumann BCs (right)}
  \label{fig:sbc1}
\end{figure}

\begin{figure}[h!]
  \centering
  \includegraphics[width=0.4\linewidth]{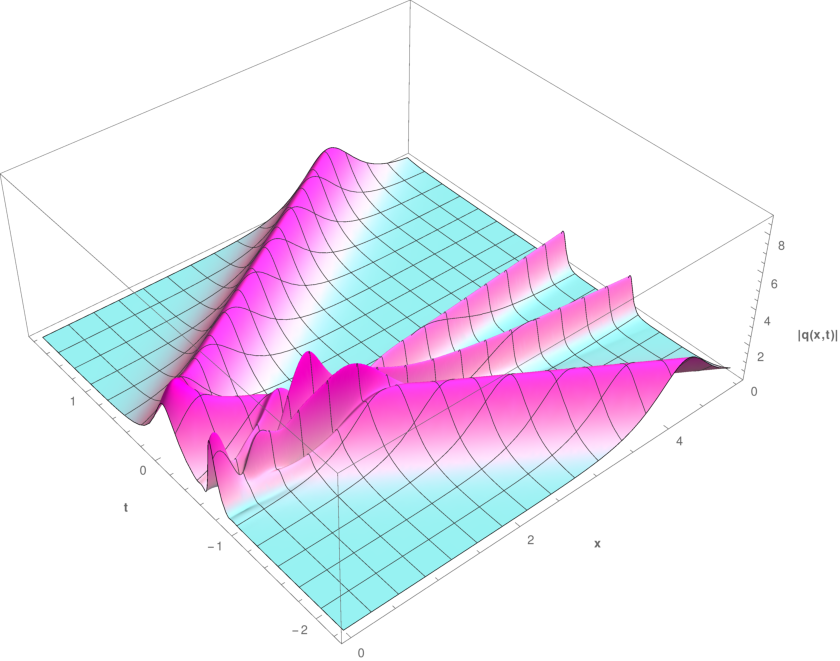} \hspace{0.2cm}
  \includegraphics[scale=0.4]{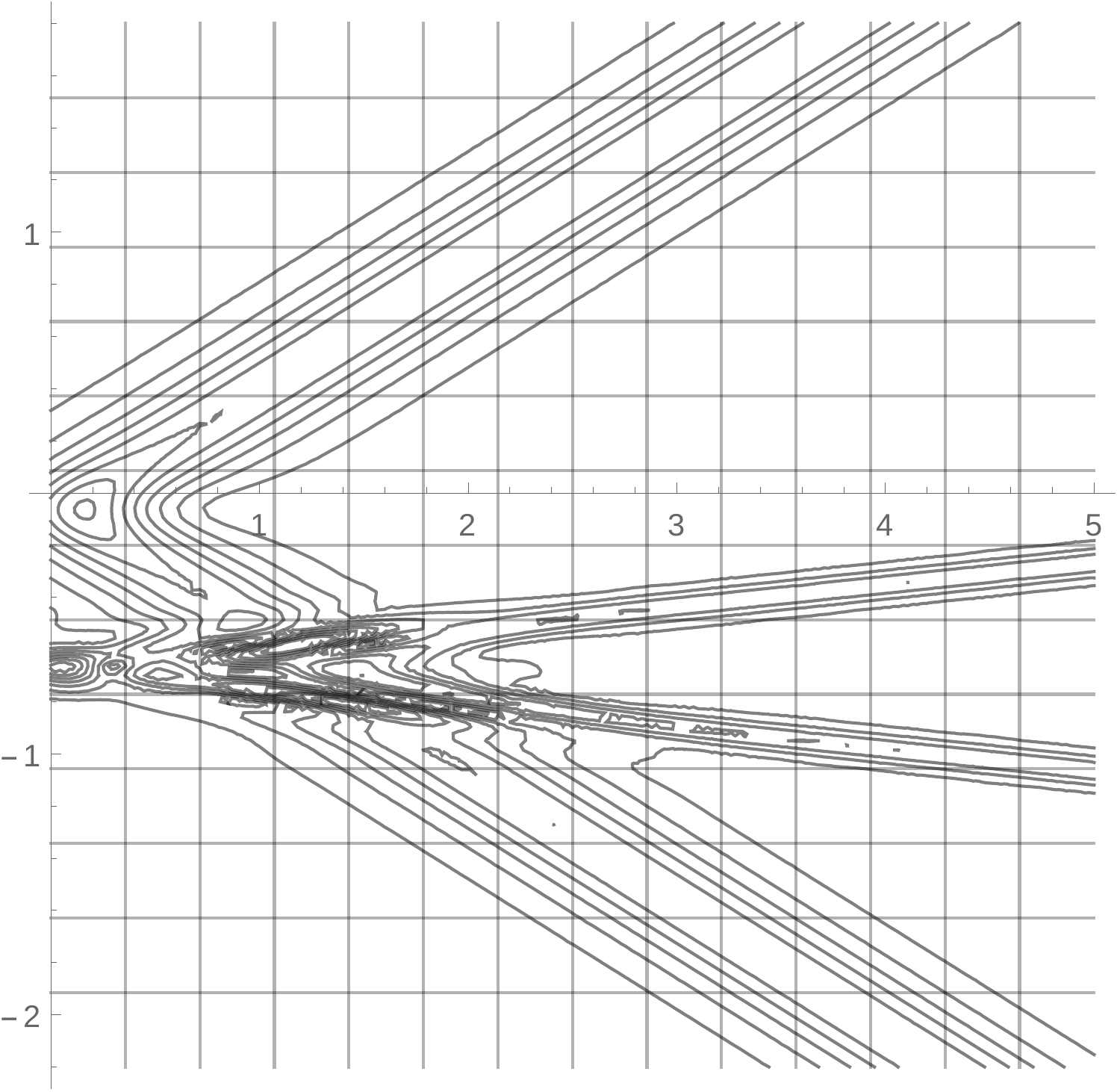}
  \caption{Two-soliton solution satisfying the Robin BCs ($\alpha=3$)}
  \label{fig:sbc}
\end{figure}

It is straightforward to apply Prop.~\ref{prop:NSS}. The zero seed solution implies that the special solution $\psi_j(\lambda_j)$, $j=1,\dots,N$, is in the form 
\begin{equation}
  \label{eq:genepsi}
  \psi_j(\lambda_j) = \bma \mu_i \\ \nu_i\ema =e^{\left(-i\lambda_jx -2i\lambda_j^2t\right)\, \sigma_3} \,  \bma u_j \\ v_j\ema\,,
\end{equation}
where $\lambda_j$ is a complex parameter. Here $ (u_j, v_j)^\intercal$ is a constant vector. Obviously,
\begin{equation}
    \widehat{\psi}_j (\widehat{\lambda}_j)= K(-\lambda_j)\,\psi_j(-\lambda_j)\,,
\end{equation}
is the paired special solution of $\psi_j(\lambda_j)$. Now having the data $\{\psi_j(\lambda_j),\widehat{\psi}_j(\widehat{\lambda}_j)\}$, we are ready to compute the $N$-soliton solutions of the NLS equation on the half-line. Two solitons interacting with the boundary are illustrated in Fig.~\ref{fig:sbc1} and \ref{fig:sbc}.   

\subsection{Boundary-bound solitons}
Static soliton solutions arise as the special solutions $\psi_j(\lambda_j)$ having pure imaginary  parameters $\lambda_j$. Since $\bar{\lambda}_j= -\lambda_j$, each $\psi_j$ needs to be paired with itself to eventually make the boundary constraint preserved  \eqref{eq:dbm1} in the boundary-dressing process. 
\begin{proposition}[Boundary-bound solitons]\label{prop:51} Assume that, associated with $N$ distinct pure imaginary parameters $\lambda_j =  i\kappa_j$, $\kappa_j \in \RR$,  there exist $N$ special solutions $\psi_j(\lambda_j)$, $j=1,\dots,N$, of the undressed Lax system \eqref{eq:laxp} with zero seed solution, in the forms
\begin{equation}
  \psi_j(\lambda_j)=\psi_j(i\kappa_j) = \bma \mu_i \\ \nu_i\ema =e^{\left(\kappa_jx +2i\kappa_j^2t\right)\, \sigma_3} \,  \bma u_j \\ 1\ema\,, \quad x\geq 0 \,,
  \end{equation}
such that  $\kappa_j$ satisfies
\begin{equation}
  \label{eq:ad51}
  f_\alpha(i\kappa_j) = \frac{\alpha+2\kappa_j}{\alpha-2\kappa_j} < 0\,,
\end{equation}
with   $\alpha$ being a real parameter and $u_j$ being defined as
\begin{equation}
  \label{eq:ujsm}
   u_j = \sqrt{-  f_\alpha(i\kappa_j)}^{(-1)^{N}}\,,
\end{equation}
then the so-construction $q[N]$ corresponds to a static $N$-soliton solution satisfying
\begin{equation}
  (q[N]_x- \alpha q[N])\rvert_{x=0} = 0\,.
\end{equation}
\end{proposition}
\prf
The proof is split   into two cases: when $N$ is odd and when $N$ is even. Moreover, for simplicity, we only consider the cases where $N=1$ and $N=2$ since the odd  $N$  and the  even  $N$ cases can be understood as the generalizations.

{\noindent  N=1:} since the zero seed solutions is imposed, one has $V[0] = -2i\lambda^2\sigma_3$. 
The one-step DT involves a dressing matrix $D[1]$ constructed from a special solution $\psi_1(\lambda_1)$. Using the identity $(iaI \pm 2\lambda \sigma_3) = (ia \mp 2\lambda) K(\pm \lambda)$, one can show if $D[1]$ obeys 
\begin{equation}
  \label{eq:511}
  (iaI-2\lambda \sigma_3)D[1](\lambda)\rvert_{x=0} =D[1](-\lambda)(iaI+2\lambda \sigma_3)\rvert_{x=0}\,,
\end{equation}
then the dressed $V[1] = D[1]_tD[1]^{-1}+D[1]V[0]D[1]^{-1}$ satisfies the boundary constraints \eqref{eq:dbm1}. It remains to prove the relation \eqref{eq:511}. Knowing that $D[1](\lambda)$ is a matrix polynomial of degree $1$  in $\lambda$, the {\it l.h.s.} and {\it r.h.s.} of \eqref{eq:511} are thus polynomials of degree $2$. Denote them by
\begin{align}
  L(\lambda)=&(iaI-2\lambda \sigma_3)D[1](\lambda) = \lambda^2L_0+\lambda^1L_1+L_2\,,\\
    R(\lambda)=&D[1](-\lambda)(iaI+2\lambda \sigma_3)= \lambda^2R_0+\lambda^1R_1+R_2\,. 
\end{align}
Clearly, $L_0=R_0$, $L_2=R_2$. This explains the ``presumed" form \eqref{eq:511}: $D[1](\lambda)$ is of odd degree in $\lambda$, thus \eqref{eq:511} ensures that the leading and zero-degree terms of the both sides are equal. The equality  \eqref{eq:511} holds, if
\begin{equation}
  \label{eq:5211}
  K(\lambda_1)\psi_1\rvert_{x=0} =\sigma_2 \bar{\psi}_1\rvert_{x=0} \,. 
\end{equation}
In fact, let $\varphi_1 = \sigma_2 \bar{\psi}_1$, $D[1](-\lambda_1)\varphi_1 =0$. One can show that
\begin{equation}
  L(\lambda)\rvert_{\lambda=\lambda_1}\psi_1=0\,,\quad   R(\lambda)\rvert_{\lambda=\lambda_1}\psi_1=0\,,\quad   L(\lambda)\rvert_{\lambda=-\lambda_1}\varphi_1=0\,, \quad  R(\lambda)\rvert_{\lambda=-\lambda_1}\varphi_1=0\,,
\end{equation}
meaning that  $L(\lambda)$ and $R(\lambda)$ share the same zeros and the associated kernel vectors, thus $L(\lambda) = R(\lambda)$. Moreover, the  constraint \eqref{eq:5211} imposes
\begin{equation}
 f_\alpha(i\kappa_1)u_1= -i \bar{v}_1\,,\quad  v_1= i \bar{u}_1\,,
\end{equation}
where $u_1,v_1$ are elements of the constant vector appearing in the general expression of $\psi_j$ \eqref{eq:genepsi}. The above constraints lead to 
\begin{equation}
  \label{eq:ad52}
  f_\alpha(i\kappa_1)|u_1|^2 = - |v_1|^2\,.
\end{equation}
Because of the requirement that $ f_\alpha(i\kappa_1)<0$, and without loss of generality, letting $|v_1|^2=1$ and $u_1$ be real, then  $u_1 = 1/\sqrt{-f_\alpha(i\kappa_1)}$. One recovers the statements \eqref{eq:ujsm} for $N=1$. Note that the assumption \eqref{eq:511} is needed  for any odd $ N$, this impose similar constraints on $u_j,v_j$ as shown in \eqref{eq:ad52}. 

{\noindent  N=2:} similarly, one needs the following identity
\begin{equation}
  \label{eq:5121}
  (iaI-2\lambda \sigma_3)D[2](\lambda)\rvert_{x=0} =D[2](-\lambda)(iaI-2\lambda \sigma_3)\rvert_{x=0}\,. \end{equation}
Then the dressed $V[2] = D[2]_tD[2]^{-1}+D[2]V[0]D[2]^{-1}$ satisfies the boundary constraints \eqref{eq:dbm1}. As previously explained, this identity ensures the equality of the leading and zero-degree terms of the both sides of \eqref{eq:5121}. The relation holds if 
\begin{equation}
  K(-\lambda_j)\psi_j\rvert_{x=0} = \sigma_2 \bar{\psi_j}\rvert_{x=0}\,,\quad j=1,2\,.
\end{equation}
In compononts, it reads
\begin{equation}
    f_\alpha(-i\kappa_j)|u_j|^2 =\frac{1}{f_\alpha(i\kappa_j)}|u_j|^2 = - |v_j|^2\,,\quad j=1,2.
\end{equation}
Again let  $|v_j|^2=1$ and $u_j$ be real, one obtains  \eqref{eq:ujsm} for $N=2$. This condition is true for any even $N$. 
\finprf
\begin{rmk}
 The requirement  $ f_\alpha(i\kappa_j)<0$  excludes  the Dirichlet BCs, \ie $\alpha\to \infty$, for the boundary-bound solitons.  
\end{rmk}

\begin{figure}[h!]
  \centering
  \includegraphics[width=0.4\linewidth]{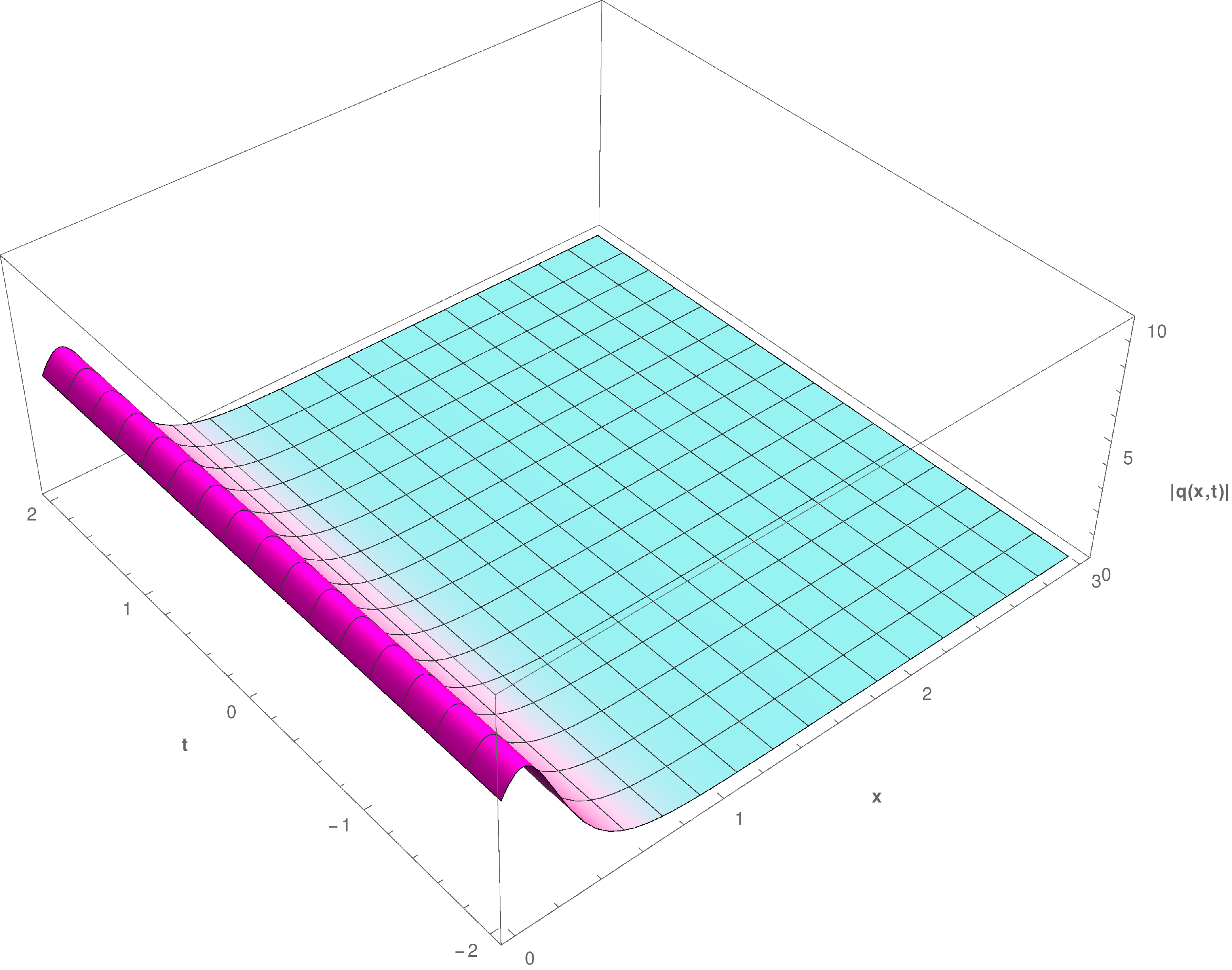} \hspace{.4cm}
  \includegraphics[width=0.4\linewidth]{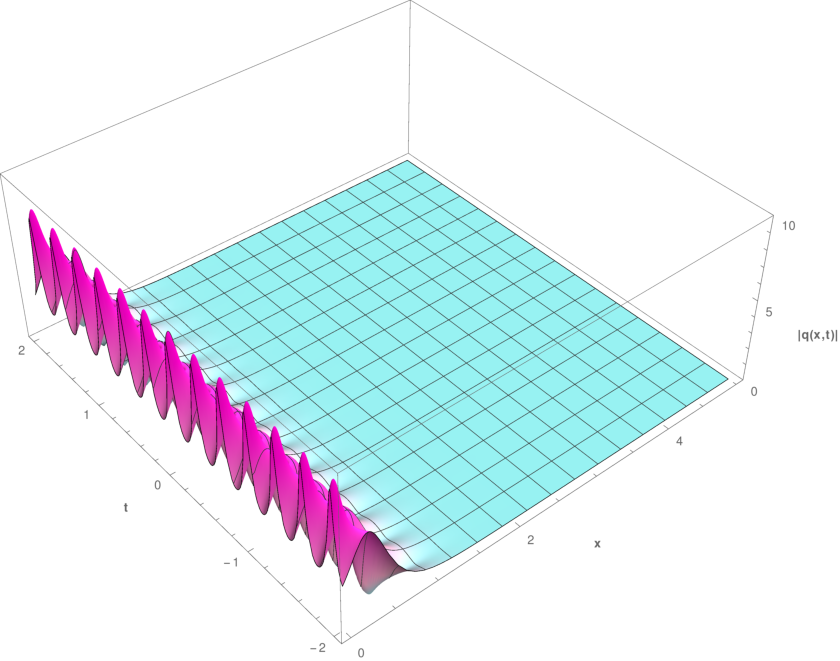}
  \caption{Boundary-bound solitons under the Robin BCs ($\alpha=3$): the magnitude (norm) is constant along the boundary for the one-soliton case (left); an interference phenomenon takes place for a doubly-boundary-bound soliton (right)}
  \label{fig:sbc3}
\end{figure}

\begin{figure}[h!]
  \centering
  \includegraphics[width=0.4\linewidth]{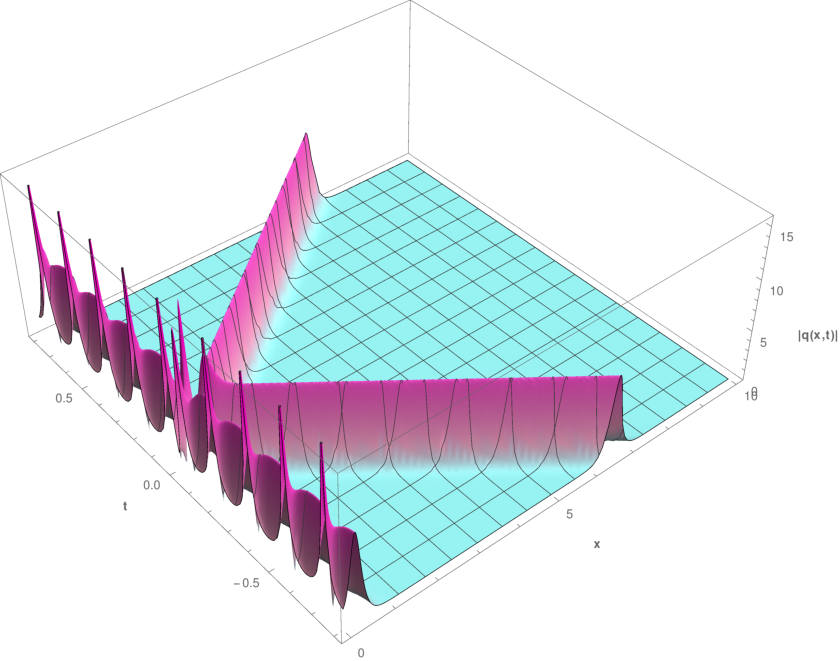} \hspace{0.2cm}
  \includegraphics[scale=0.4]{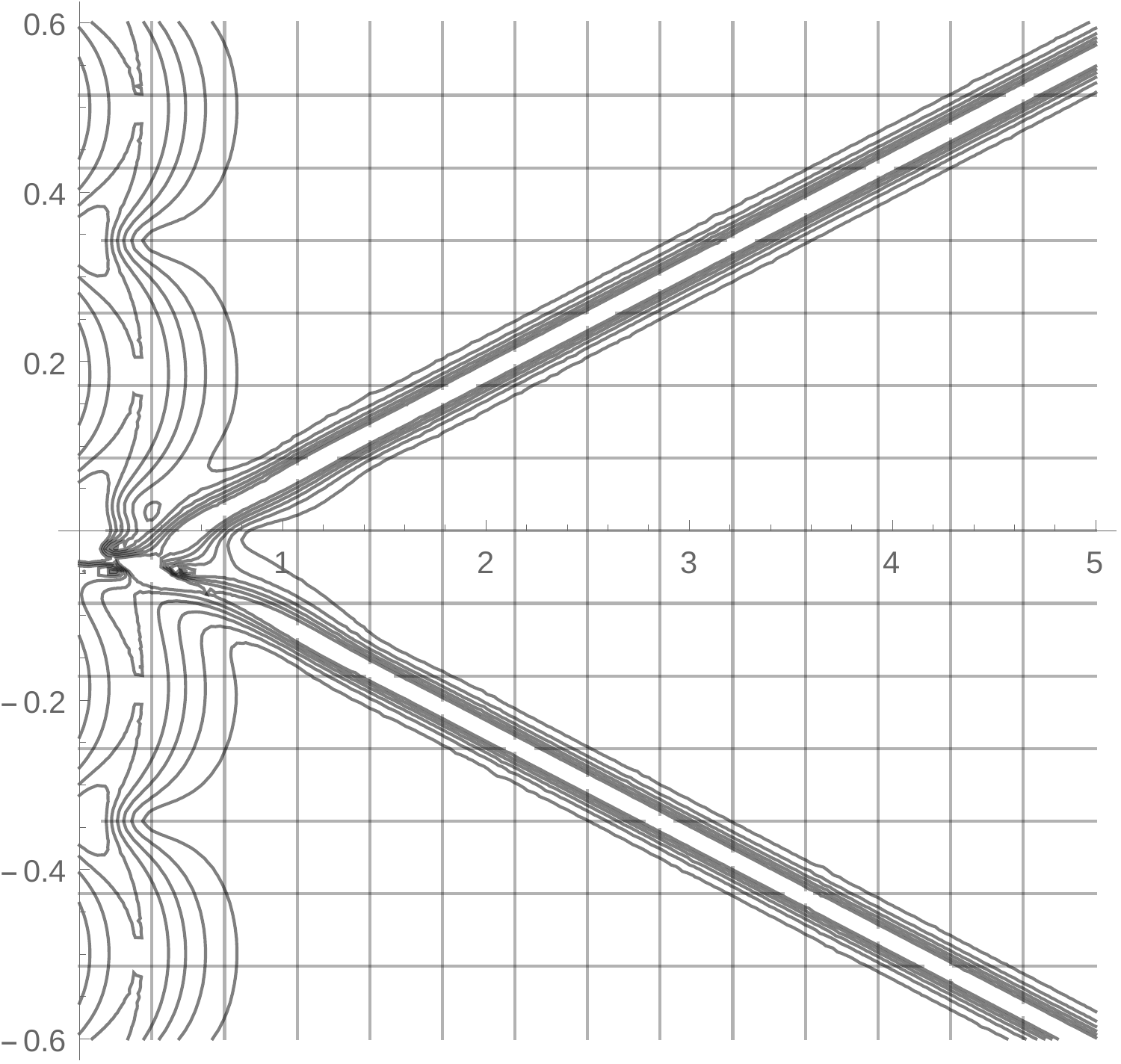}
  \caption{Interaction between a  (moving) soliton and a doubly-boundary-bound soliton under the Robin BCs ($\alpha=3$)}
  \label{fig:sbc4}
\end{figure}

Following the above proposition, one can easily compute static solitons bounded to the boundary under the Robin BCs.  When there are multi-static solitons bounded to the bounadry,  the interference phenomena take place (see Fig.~\ref{fig:sbc3}).  
One can dress the boundary using both the moving and static soliton data for the boundary constraint \eqref{eq:dbm1} is preserved at each step of the dressing process (see Fig.~\ref{fig:sbc4}).  Note that similar results were obtained in \cite{bion2} using the mirror-image technique.  
\section{Half-line solitons on a constant background}
\label{sec:nls2}
In this section, we are dealing with the non-zero seed solution 
\begin{equation}
  \label{eq:nzss1}
  q[0] = \rho\, e^{2i \rho^2 t}\,, \quad \rho > 0\,, 
\end{equation}
where  $\rho$ represents the constant background. 
The model is more involved than the zero seed solution case because a two-valued function  related to the spectral parameter of the special solution appears, cf.~\cite{ZS3, faddeev, AK1, AK2}. It admits self-modulated soliton solutions. 
\subsection{Analysis of the model}
We briefly recall the DT approach to the model, cf.~\cite{Manas}.  It follows from a gauge transformation $\phi \to D_{\rho}\phi$ to the undressed Lax system \eqref{eq:laxp}  with 
\begin{equation}
  D_{\rho} =\bma 1 & 0 \\ 0 & e^{2i\rho^2 t}  \ema\,,
\end{equation}
that the Lax pair is transformed to two constant matrices
\begin{equation}
 {\mathcal U}= -i\lambda\,\sigma_3+\bma 0 & \rho \\ -\rho & 0 \ema\,,\quad{\mathcal V} = -i\,{\mathcal U}^2+2\lambda\,{\mathcal U}+i(\lambda^2I+2\rho ^2)\,,
\end{equation}
The eigenvalues of  $\mathcal U$  are  $\pm i \xi$ with $\xi$ satisfying
\begin{equation}\label{eq:tsrs1}
\xi^2 = \lambda^2-\rho^2    \,,
  \end{equation}
  and those of  $\mathcal V$ are $i (\rho^2 \pm 2 \lambda\,\xi) $. Here $\xi$ can be seen as a the two-valued function of $\lambda$.
 The matrices  ${ \mathcal U}, { \mathcal V}$ can be simultaneously  diagonalized following
\begin{equation}
  \label{eq:laxp11}
  M^{-1}{\mathcal U}M =i\xi\sigma_3\,, \quad  M^{-1}{\mathcal V}M =i (\rho^2I+2\lambda\,\xi\sigma_3 )\,,\end{equation}
where \begin{equation}
 M:= M(\lambda,\xi) = \bma i(  \lambda -\xi)/\rho& 
1 \\ 1& i(  \lambda -\xi)/\rho
\ema\,.
\end{equation}
Combining the above analysis, a special solution of the undressed Lax system \eqref{eq:laxp} with the constant background seed solution \eqref{eq:nzss1} is in the form
\begin{equation}
  \label{eq:genepsi}
  \psi_j(\lambda_j,\xi_j) = \bma \mu_i \\ \nu_i\ema=D^{-1}_\rho M(\lambda_j,\xi_j)\,e^{i\left(\rho^2tI+\xi_j(x +2\lambda_jt)\sigma_3\right)} \,  \bma u_j \\ v_j\ema\,,
\end{equation}
where $\xi_j$ and  $\lambda_j$ are related by \eqref{eq:tsrs1}, and $u_j, v_j$ are constants. Although $\xi_j$ depends on $\lambda_j$, we put the explicit dependence of $\xi_j$ because the sign is important in later determination of soliton solutions on the half-line.  
Clearly, inserting $ \mu_i, \nu_i$, $j=1,\dots,N$, into \eqref{eq:solqn} gives rise to  $N$-soliton solutions of the model.
\subsection{Soliton solutions on the half-line under the Neumann BCs}
One can easily check that  the seed solution \eqref{eq:nzss1} is subject to the Neumann BCs and 
\begin{equation}
\label{eq:csnzs}
  \sigma_3\, {V}[0](-\lambda)\rvert_{x=0} =   {V}[0](\lambda)\rvert_{x=0} \,\sigma_3\,,
\end{equation}
with  $ {V}[0]$ being the  $t$-part of the undressed Lax pair.  Now we are looking to dress the Lax pair by preserving the boundary constraint \eqref{eq:csnzs}. 

Consider the case $\lambda_j\neq -\bar{\lambda}_j$. Given $\psi_j(\lambda_j,\xi_j)$ in the form \eqref{eq:genepsi}, let the paired special solution be in the form
\begin{equation}
  \widehat{\psi}_j(\widehat{\lambda}_j,\widehat{\xi}_j)= \bma \widehat{\mu}_i \\ \widehat{\nu}_i\ema=D^{-1}_\rho M(-\lambda_j,-\xi_j)\,e^{i\left(\rho^2tI-\xi_j(x -2\lambda_jt)\sigma_3\right)} \,  \bma -u_j \\ v_j\ema\,. 
\end{equation}
Using the identity
$
  \sigma_3 M(\lambda_j,\xi_j) = -  M(-\lambda_j,-\xi_j) \sigma_3$, 
one can verify that
\begin{equation}
  \widehat{\psi}_j(\widehat{\lambda}_j,\widehat{\xi}_j) \rvert_{x=0} = \sigma_3 \, \psi_j(\lambda_j,\xi_j) \rvert_{x=0}\,.
\end{equation}
One can see that the paired special solution requires not only $\widehat{\lambda}_j =-\lambda_j$ but also  $\widehat{\xi}_j =-\xi_j$.  It is straightforward to apply Prop.~\ref{prop:NSS} to obtain  $N$-soliton solutions on the half-line under the Neumann BCs.  
\begin{figure}[h!]
  \centering
  \includegraphics[width=0.4\linewidth]{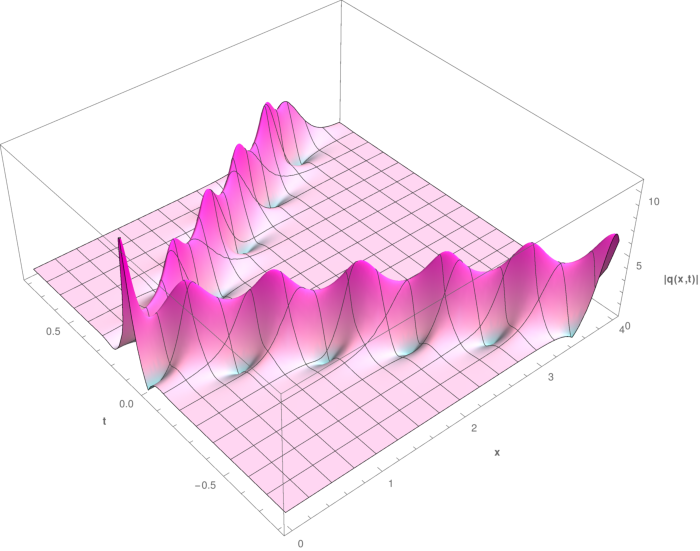} \hspace{.4cm}
  \includegraphics[width=0.4\linewidth]{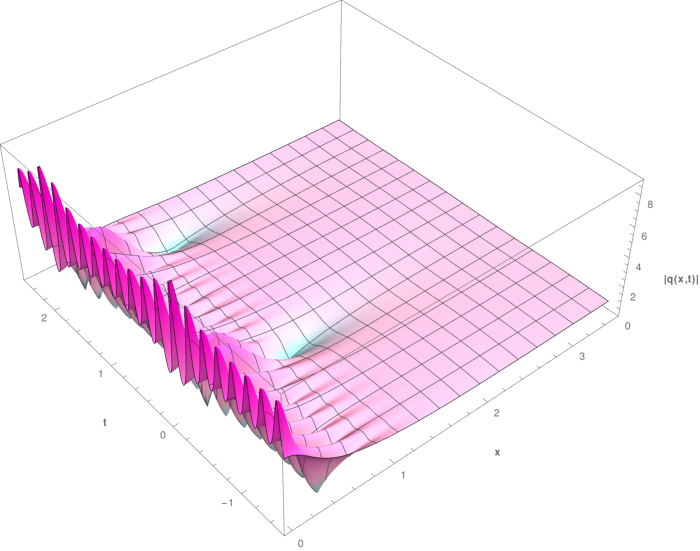}
  \caption{A self-modulated soliton on a constant background interacting with the boundary  (left) and a doubly-boundary-bound soliton (right)}
  \label{fig:sbc7}
\end{figure}

\begin{figure}[h!]
  \centering
  \includegraphics[width=0.4\linewidth]{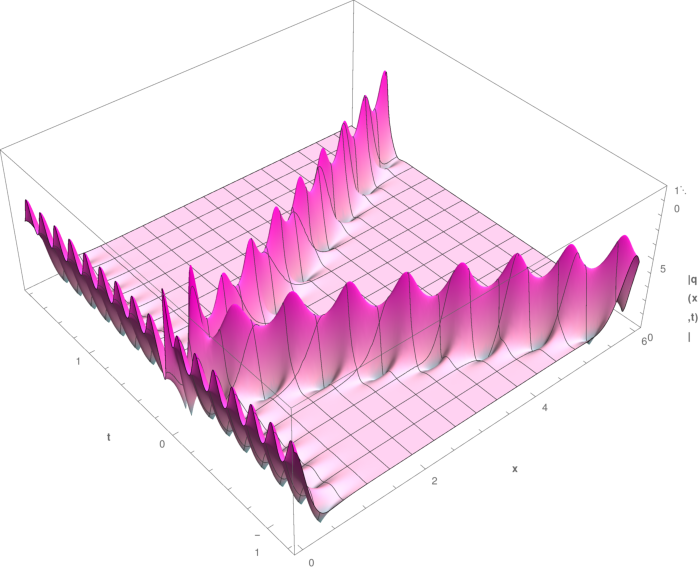} \hspace{0.2cm}
  \includegraphics[scale=0.4]{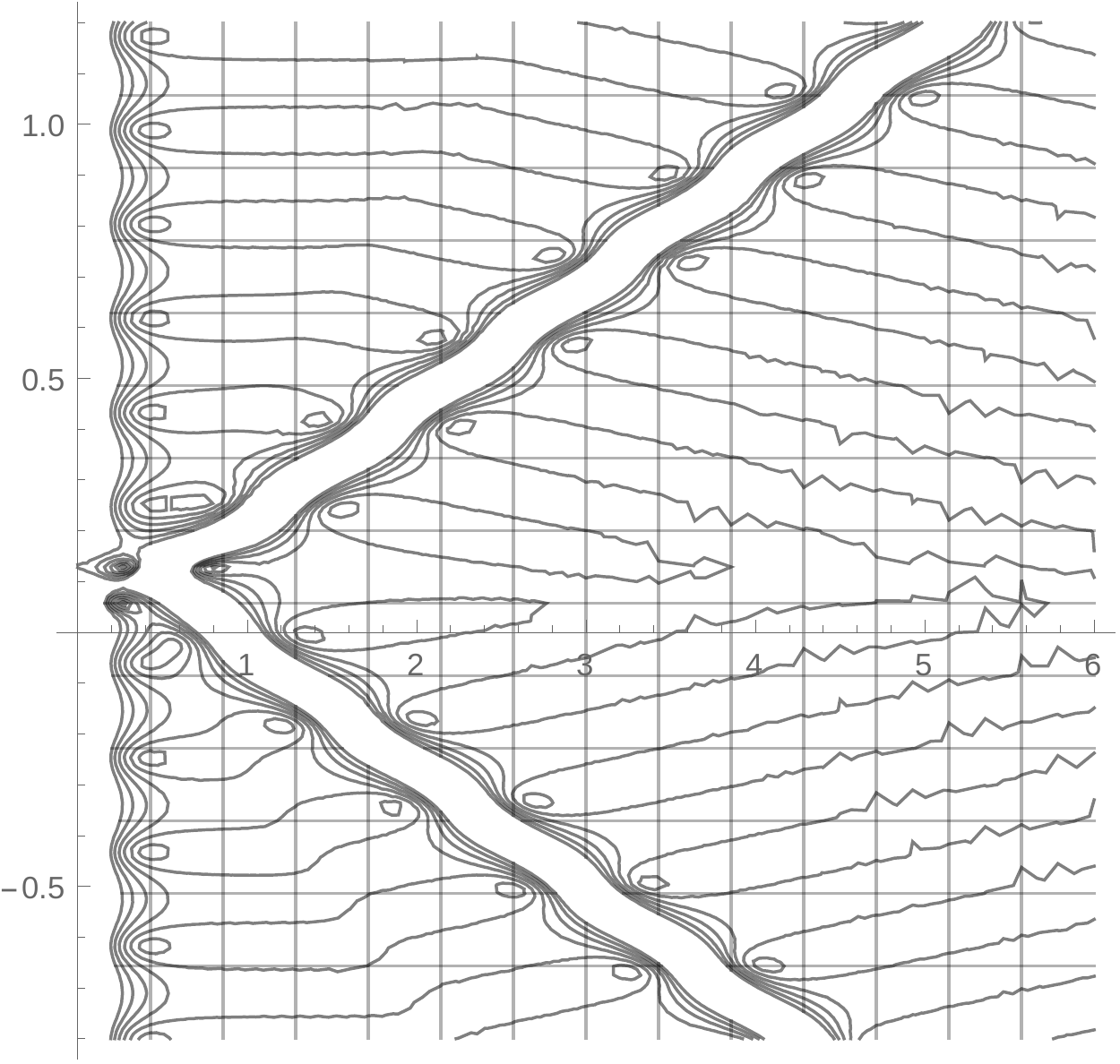}
  \caption{Interaction between a  (moving) soliton and a boundary-bound soliton under the Neumann BCs on a constant background}
  \label{fig:sbc8}
\end{figure}

For static solitons ($\lambda_j = -\bar{\lambda}_j$), one can used the similar ideas as presented in Prop.~\ref{prop:51}. By letting $u_j=v_j=1$, one can show the boundary constraint \eqref{eq:csnzs} is preserved at each step of the dressing. Examples of half-line self-modulated solitons are illustrated in Fig.~\ref{fig:sbc7}.  
Moreover, one can combine the (moving) soliton and boundary-bound solitons together (see Fig.~\ref{fig:sbc8}).  
\section{A ``space-evolution'' interpretation}
\label{sec:7}
The IST is an analytic method for solving initial-value problems for integrable PDEs with the space-time domain restricted to $x \in \RR$, $t\geq t_0$. It is essentially made of three steps: first, the direct scattering process where the initial conditions at $t=t_0$ are transformed into scattering data using the $x$-part of the Lax pair; second, the {\em  time-evolution process} where the scattering data {evolve linearly in time} according to the $t$-part of the Lax pair; and last, the inverse problem where the scattering data are put into the reconstruction formulae to recover solutions of the integrable PDEs.

In contrast to the usual IST, the unified transform approach to half-line problems is to restrict the space-time domain to $x \geq 0$, $t\geq t_0$. Then,  in the direct scattering process one encodes both the initial and boundary conditions into the scattering data. However, it is a hard problem to solve the inverse problem even as for a rather simple situation such as the NLS equation on the half-line where exact solutions do exist. 

In order to overcome this problem and to fit our approach to exact solutions on the half-line into the IST scheme, one needs to extend the space-time domain into $x\ge0$, $t\in \RR$. This can be compared with the usual IST, and in turn, a boundary-value problem is defined where the ``initial boundary profile" is imposed by the BCs at $x=0$. It turns out that the boundary-value problem can be solved using a space-evolution process where scattering data are determined by the $t$-part of the Lax pair at $x=0$ and  evolve linearly in space for $x\geq 0$.

To make the statements precise, we can perform the direct scattering for the $t$-part of the Lax pair $V$ at $x=0$. For simplify, we only consider the zero seed solution case. This imposes the vanishing asymptotic conditions under which the NLS field $q$ and its $x$ derivatives vanish as $t\to \pm \infty$. Then one can have the Jost solutions
\begin{equation}
\lim_{t\to \pm\infty} \phi_\pm(0,t;\lambda) = e^{-2i\lambda^2t\sigma_3}\, .   
\end{equation}

\begin{figure}[th]
  \centering
  \begin{tikzpicture}[scale=1]
    \def\l{3}%
    \coordinate (p1) at (-\l,0);
    \coordinate (p2) at (\l,0);
    \coordinate (q1) at (0,-\l);
    \coordinate (q2) at (0,\l);
    \draw[->] (p1)  -- (p2);
    \draw[->] (q1)  -- (q2) ;
    \node at (\l*0.85, \l*0.85) {$(1)$};
    \node at (-\l*0.85, \l*0.85) {$(2)$};
    \node at (\l*0.85, -\l*0.85) {$(4)$};
    \node at (-\l*0.85, -\l*0.85) {$(3)$};
    \draw[<->]  [ dashed] (\l/2,\l/2) node [above right] {$\bar{\lambda}_j$} --  (- \l/2, - \l/2) node [above left] {$\bar{\widehat{\lambda}}_{j}$};
    \draw[<->]  [ dashed] (\l/2, -\l/2) node [above right] {$\lambda_j$} --  (- \l/2,  \l/2) node [above left] {$\widehat{\lambda}_{j} $};
  \end{tikzpicture} 
  \caption{Zeros of the scattering function ${a}^{(24)}(\lambda)$ in the presence of Robin BCs} \label{fig:51011}
\end{figure}
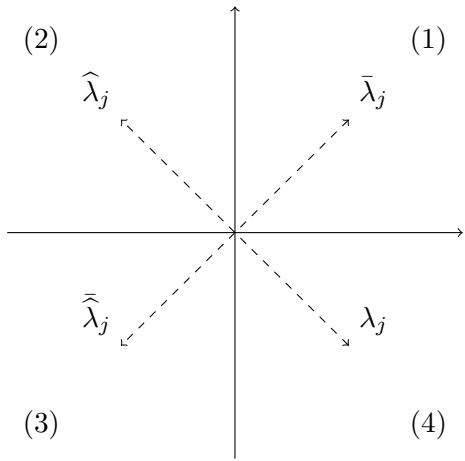

Due to the $\lambda^2$-dependence of  the spectral parameter, the analytical domain of Jost solutions can be naturally split into four quadrants, which leads to a ``time" monodromy matrix   $S(k)$ in the form \begin{equation}
\label{eq:41rep_T}
S(\lambda) =  \begin{pmatrix}
a^{(24)}(\lambda)&\bar{b}^{(13)}(\lambda)\\
{b}^{(24)}(\bar{\lambda})&\bar{a}^{(13)}(\bar{\lambda})
\end{pmatrix} \,. 
\end{equation}
Here the subscript of ${a}^{(24)}(\lambda)$ means that the scattering function ${a}^{(24)}(\lambda)$ can be analytically continued to the union of the quadrants $(2)$ and $(4)$ (see Fig.~\ref{fig:51011} for the distribution of the quadrants). This applies also to other scattering functions. 

Apparently,  the direct scattering of $V$ at $x=0$  differs from the usual IST only by the use of potential. Here the potential is  $Q_T$ as $V$ can be written as
\begin{equation}
  V = -2i\lambda^2\sigma_3+Q_T\,,\quad Q_T= 2\lambda \, Q -iQ_x\,\sigma_3-i Q^2\,\sigma_3\,.
\end{equation}
This switches the roles of initial and  boundary conditions: instead of characterizing initial profile, the BCs are first considered and  encoded into $S(k)$. Then following the asymptotic conditions imposed to the NLS field $q$ as $t\to \pm\infty$, one can easily show that the scattering data evolve linearly in $x$ as $S( \lambda)$ obeys  
\begin{equation}
  \label{eq:41spacee}
  \frac{\partial S( k)}{\partial x} = - i\lambda [\sigma_3, S( k)]\,.
\end{equation}
The Jost solutions can be put into certain $(x,t)$-dependent Riemann-Hilbert problem, which eventually lead to soliton solutions of NLS with zeros of   ${a}^{(24)}(\lambda)$ appearing in the union of the quadrants $(2)$ and $(4)$. 

Having the above space-evolution process in mind, we are ready to implement the Robin BCs into the system. Since $V$ is required to obey the boundary constraint \eqref{eq:dbm1},  an additional constraint on $S$ appears
\begin{equation}\label{eq:42sym1}
  S(-\lambda) = K(-\lambda)\,S(\lambda)\,K(\lambda)\,,
\end{equation}
with the boundary matrix $K(\lambda)$ defined in \eqref{eq:formkf}. Consequently, the Robin BCs implies that if $\lambda_j$ is a zero of ${a}^{(24)}(\lambda)$, so does $-\lambda_j$ because
\begin{equation}
  {a}^{(24)}(-\lambda) ={a}^{(24)}(\lambda)\,.
\end{equation}
The associated norming constants\footnote{The norming constants can be understood as ratio of $u_j$ and $v_j$ appearing in the special solutions in the DT process.}  are related by
\begin{equation}
  b^{(24)}(-\lambda_j) = f_a(-\lambda)   b^{(24)}(\lambda_j)\,.
\end{equation}
Therefore, the paired zeros of ${a}^{(24)}(\lambda)$ (see Fig.~\ref{fig:51011}) and the relation between the paired norming constants give the underlying reason of the paired special solutions $\{\psi_j(\lambda_j), \widehat{\psi}_j(\widehat{\lambda}_j)\}$ in boundary-dressing process in Prop.~\ref{prop:NSS}. Note that the  relation \eqref{eq:42sym1} is in contrast to the mirror-image technique where the pairing of zeros of the scattering function is related by spectra $\{\lambda_j, -\bar{\lambda}_j\}$, cf.~\cite[Equation (2.33)]{biondini2009solitons}.

\section{Conclusions}
By carefully reviewing the integrable BCs for the NLS equation, we provide a direct method for computing exact solutions of the focusing NLS equation on the half-line under the Robin BCs.  The method is lying on dressing the integrable boundary constraints by appropriate pairing of special solutions in the Darboux-dressing process. Two classes of seed solutions are considered, which lead to usual (bright) solitons on the half-line  and self-modulated solitons on the half-line  respectively.   In particular, the boundary-bound solitons are computed in both cases. The method has the advantage that it is simple and direct. It admits a natural IST interpretation as evolution in space of the integrable boundary data. 

It is believed that the boundary-dressing approach can be applied to a wide range of problems where the integrable boundary structures exist. As to the NLS case, one can, for instance, compute half-line dark solitons which correspond to exact solutions of the defocusing NLS equation  on the half-line\footnote{The work is in progress.}.  Applications of the boundary-dressing method to computing exact half-line solutions of the sine-Gordon equation was recently obtained  \cite{ZCZ}.  Other extensions of the method can be related to dressing the boundary on a star-graph \cite{Cau} where similar boundary constraints appear, or to tackle integrable PDEs on an interval where the algebraic-geometric integration technique may be involved \cite{N1, bel1}.

\section*{Acknowledgments}
The author is supported by NSFC (No.11601312) and Shanghai Young Eastern Scholar program (2016-2019).


\begin{thebibliography}{99}

\bibitem{kdv}
  Gardner CS, Greene JM, Kruskal MD, Miura RM,
  Method for solving the Korteweg-de Vries equation.
  {\em Physical review letters}, 19(19), pp. 1095, (1967).


\bibitem{G1}
  Gardner CS,
  Korteweg-de Vries Equation and Generalizations. IV. The
  Korteweg-de Vries Equation as a Hamiltonian System.
  {\em Journal of Mathematical Physics}, 12(8), pp. 1548--1551, (1971).

\bibitem{f1}
  Zakharov VE, Faddeev LD,
  Korteweg-de Vries equation: A completely integrable Hamiltonian system.
  {\em Functional analysis and its applications}, 5(4), pp. 280--287, (1971).

\bibitem{sklyanin1987boundary}
{Sklyanin EK},
\newblock Boundary conditions for integrable equations.
\newblock {\em Functional Analysis and its Applications}, 21(2), pp. 164--166, (1987).

\bibitem{faddeev}
Faddeev LD, Takhtajan LA,
\newblock {\em {Hamiltonian Methods in the Theory of Solitons}}.
\newblock Springer Science \& Business Media, (2007).


\bibitem{SKBC}
{Sklyanin EK},
\newblock Boundary conditions for integrable equation.
\newblock {\em Functional Analysis and its Applications}, 21(2), pp. 164--166, (1987).

 
\bibitem{Cherednik1}
  Cherednik IV,
  Factorizing particles on a half-line and root systems.
 {\em  Theoretical and Mathematical Physics}, 61(1), pp. 977--983, (1984).

\bibitem{AKNS} Ablowitz MJ, Kaup DJ, Newell AC,  Segur H,
  The Inverse Scattering Transform-Fourier Analysis for Nonlinear Problems.
  {\em Studies in Applied Mathematics}, 53(4), pp. 249-315, (1974).

\bibitem{fokas1997unified}
Fokas AS,
\newblock {A unified transform method for solving linear and certain nonlinear  PDEs}. 
\newblock {\em Proceedings of the Royal Society of London. Series A:
  Mathematical, Physical and Engineering Sciences}, 453(1962), pp. 1411--1443, (1997).

\bibitem{fokas2002integrable}
Fokas AS,
\newblock {Integrable nonlinear evolution equations on the half-line}.
\newblock {\em Communications in mathematical physics}, 230(1), pp. 1--39, (2002).

\bibitem{fokas2008}Fokas AS,
  {\em A unified approach to boundary value problems}. 
  SIAM, (2008).

\bibitem{ZS2} Zakharov VE, Shabat AB,
  Exact theory of two-dimensional self-focusing and one-dimensional self-modulation of waves in non-linear media.
  {\em Soviet Physics, JETP}, 34(1), pp. 62--9, (1972).

\bibitem{APT} Ablowitz MJ, Prinari B,  Trubatch AD,
  {\em Discrete and continuous nonlinear Schr\"odinger systems}.
  Cambridge University Press, (2004).

\bibitem{ZS3} Zakharov VE, Shabat AB,
  Interaction between solitons in a stable medium.
  {\em Soviet Physics, JETP}, 37(5), pp. 823--828,  (1973).

\bibitem{AK1} Asano N, Kato Y,
  Non‐self‐adjoint Zakharov–Shabat operator with a potential of the finite asymptotic values. I. Direct spectral and scattering problems.
  {\em Journal of Mathematical Physics}, 22(12), pp. 2780--2793, (1981).
\bibitem{AK2} Asano N, Kato Y,
  Non-self‐adjoint Zakharov–Shabat operator with a potential of the finite asymptotic values. II. Inverse problem.
  {\em Journal of mathematical physics}, 25(3), pp. 570-588, (1984).
  
\bibitem{fokas1989initial}
Fokas AS,
\newblock {An initial-boundary value problem for the nonlinear Schr{\"o}dinger  equation}.
\newblock {\em Physica D: Nonlinear Phenomena}, 35(1), pp. 167--185, (1989).

\bibitem{BT1} Bikbaev RF,  Tarasov VO, 
  Initial boundary value problem for the nonlinear Schr\"odinger equation.
 {\em Journal of Physics A: Mathematical and Theoretical}, 24(11), pp. 2507--2516, (1991).

\bibitem{Tarasov} Tarasov VO,
  The integrable initial-boundary value problem on a semiline: nonlinear Schrodinger and sine-Gordon equations.
  {\em Inverse Problems}, 7(3), pp. 435, (1991).

\bibitem{HH1}
  Habibullin IT,
  The B\"acklund transformation and integrable initial boundary value problems.
  {\em Matematicheskie Zametki}, 49(4), pp.130--137, (1991).

\bibitem{fokas1996linearization}
Fokas AS, Its AR,
\newblock {The linearization of the initial-boundary value problem of the
  nonlinear Schr\"odinger equation}.
\newblock {\em SIAM Journal on Mathematical Analysis}, 27(3), pp. 738--764, (1996).


\bibitem{fokas2005nonlinear}
Fokas AS, Its AR, Sung LY,
\newblock {The nonlinear Schr{\"o}dinger equation on the half-line}.
\newblock {\em Nonlinearity}, 18(4), pp. 1771, (2005).

  
\bibitem{biondini2009solitons}
  Biondini G, Hwang G,
\newblock {Solitons, boundary value problems and a nonlinear method of images}.
\newblock {\em Journal of Physics A: Mathematical and Theoretical},
  42(20), pp. 205--207, (2009).

\bibitem{CZ}
{{Caudrelier} V, {Zhang} C,}
\newblock {The vector nonlinear Schr{\"o}dinger equation on the half-line}.
\newblock {\em Journal of Physics A: Mathematical and Theoretical},
45(10), pp. 105201, (2012).

  
\bibitem{bion2} Biondini G, Bui A,
  On the nonlinear Schr\"odinger equation on the half line with homogeneous Robin boundary conditions.
  {\em Studies in Applied Mathematics}, 129(3), pp. 249--271, (2012).

\bibitem{DP1} Deift P,  Park J,
  Long-time asymptotics for solutions of the NLS equation with a delta potential and even initial data.
  {\em International Mathematics Research Notices}, 2011(24), pp. 5505--5624, (2011).

\bibitem{ZS4} Zakharov VE, Shabat AB,
  A scheme for integrating the nonlinear equations of mathematical physics by the method of the inverse scattering problem. I.
  {\em Functional analysis and its applications}, 8(3), pp. 226--235, (1974).

\bibitem{ZS5} Zakharov VE, Shabat AB,
  Integration of nonlinear equations of mathematical physics by the method of inverse scattering. II.
  {\em Functional Analysis and Its Applications}, 13(3), pp. 166--174, (1979).


\bibitem{matveev1991darboux}
Matveev VB,  Salle MA, 
\newblock {\em {Darboux transformations and solitons}}.
\newblock Springer-Verlag, (1991).

\bibitem{ACC} Avan J, Caudrelier V, Cramp\'e N,
  From Hamiltonian to zero curvature formulation for classical integrable boundary conditions. 
  {\em Journal of Physics A: Mathematical and Theoretical}, 51(30), (2018).

\bibitem{babelon}
Babelon O, Bernard D,  Talon M,
\newblock {\em {Introduction to Classical Integrable Systems}}.
\newblock {Cambridge University Press}, {2003}.


\bibitem{Manas} Manas M,
  Darboux transformations for the nonlinear Schr\"odinger equations.
{\em  Journal of Physics A: Mathematical and General}.  29(23), pp. 7721, (1996). 

\bibitem{ZCZ}
{{Zhang} C, Cheng Q, {Zhang} D-J},
\newblock {Soliton solutions of the sine-Gordon equation on the half-line}.
\newblock {\em Applied Mathematics Letters}, 86, pp. 64--69, (2018). 

\bibitem{Cau} Caudrelier V,
  On the inverse scattering method for integrable PDEs on a star graph.
  {\em Communications in Mathematical Physics}, 338(2), pp. 893--917, (2015).

\bibitem{N1} Novikov SP,
  The periodic problem for the Korteweg—de vries equation.
  {\em Functional analysis and its applications}, 8(3), pp. 236--246, (1974).

\bibitem{bel1} Belokolos ED,
  {\em Algebro-geometric approach to nonlinear integrable equations}.  Springer, (1994).	




\end{thebibliography}
\appendix
\section{Derivation of $D[N]$}
\label{ap:1}
Recall the series expansion of $D[N]$ 
\begin{equation}
  D[N] = \lambda^N +\lambda^{N-1 }\,\Sigma_1 + \lambda^{N-2 }\,\Sigma_2 \cdots+ \Sigma_N\,.
\end{equation}
The complete determination of $\Sigma_j$ relies on the identification of the kernel vectors of $D[N](\lambda)$  at its zeros $\lambda_j, \bar{\lambda}_j$
. This can be done with the help of the following lemma.
 \begin{lemma}
\label{le:11}
Let  $\psi_{j} = (\mu_j,\nu_j)^\intercal$, $j = 1\,\dots\,N$, be special solutions of the Lax system~\eqref{eq:laxp} evaluated at  $\lambda_j$. Then a set of $N$ vector functions $\varphi_j$, defined by
\begin{equation}
  \label{eq:cstc1}
    \varphi_j = c \sigma_2 \,\bar{\psi}_j =\bma -\bar{\nu}_j \\ \bar{\mu}_j \ema \,,\quad c=-i\,,\quad    \sigma_2 = \bma 0 & -i \\ i & 0 \ema\,. 
\end{equation}
 satisfy 
\begin{equation}
\label{eq:apvp}
  U(\bar{\lambda}_j)\, \varphi_j = \varphi_{j x}\,,\quad   V(\bar{\lambda}_j)\, \varphi_j = \varphi_{j t}\,. 
\end{equation}
\end{lemma}
The proportionality constant $c$ in \eqref{eq:cstc1} is irrelevant and can be replaced by any non-zero number.  It is easy to check  $\psi_j$ and $\varphi_j$ satisfy the orthogonality condition
\begin{equation}
  \varphi^\dagger_j\,\psi_j = 0\,,\quad j = 1,\cdots,N\,.
\end{equation}
The dressing matrix $D[N]$ has kernel vectors $\varphi_j$, $j=1\,\dots\,N$ at $\lambda = \bar{\lambda}_j$, thus  $\psi_j$ and  $\varphi_{j}$ give us a complete characterisation of $\Sigma_j$. 
\begin{lemma}
For $j=1,\,\dots, \, N$, the vector functions $\psi_j$, $\varphi_j$ satisfy
\begin{equation}
\label{eq:systems1}
  D[N]\big\rvert_{ \lambda= \lambda_j} \,  \psi_j = 0\,,\quad  D[N]\big\rvert_{ \lambda= \bar{\lambda}_j} \,  \varphi_j  = 0 \,.
\end{equation}
\end{lemma}
The above  system can be arranged  to a set of algebraic equations
\begin{equation}
\begin{aligned}
\left(  \lambda_1^N +\lambda_1^{N-1}\,\Sigma_1 +  \cdots + \Sigma_N\right) \, \psi_1 = 0\,, \quad &\left(  \bar{\lambda}_1^N +\bar{\lambda}_1^{N-1}\,\Sigma_1 +  \cdots + \Sigma_N\right) \, \varphi_1 = 0\,, \\
  ~~~~~~~~~~~~~~~~~~~~~~~\vdots\\
\left(  \lambda_N^N +\lambda_N^{N-1}\,\Sigma_1 +  \cdots + \Sigma_N\right) \, \psi_N = 0\,, \quad 
&\left(  \bar{\lambda}_N^N +\bar{\lambda}_N^{N-1}\,\Sigma_1 +  \cdots + \Sigma_N\right) \, \varphi_N = 0\,. 
\end{aligned}
\end{equation}
In  matrix form, one has
\begin{equation}
  \begin{aligned}
  \left(\Sigma_1\, , \cdots\,, \Sigma_N \right) \, &\bma  
\lambda^{N-1}_1\, \psi_1 & \bar{\lambda}^{N-1}_1\, \varphi_1 & \cdots & \lambda^{N-1}_N\, \psi_N       & \bar{\lambda}^{N-1}_N\, \varphi_N  \\
 \vdots & \vdots &  \vdots  & \vdots & \vdots  \\
 \psi_1 & \varphi_1 & \cdots & \psi_N       &  \varphi_N   
\ema \\
= & -  \left( \lambda_1^{N} \psi_1 \,,\bar{\lambda}_1^{N} \varphi_1\,, \cdots\,,\lambda_N^{N} \psi_N\,, \bar{\lambda}_N^{N} \varphi_N \right)  \,.
  \end{aligned}
\end{equation}
The matrix terms $\Sigma_j$ can be solved using the Cramer's rule, provided that $\psi_j$, $j=1,\dots,N$, are linearly independent. In particular, solving $\Sigma_1$ leads to the expression of $q[N]$. 
\end{document}